\documentclass[10pt,leqno]{amsart}
\usepackage{graphicx}
\usepackage{indentfirst,csquotes}

\usepackage{amssymb,amsthm,amsmath}
\usepackage{xcolor,paralist,hyperref,fancyhdr,etoolbox}

\hypersetup{ colorlinks=true, linkcolor=black, filecolor=black, urlcolor=black }

\usepackage{booktabs}
\usepackage{algorithm}
\usepackage{algpseudocode}
\usepackage{enumitem}
\usepackage{todonotes}
\usepackage[numbers]{natbib}

\newcommand{\Dhat}{\hat{D}}
\newcommand{\Dbar}{\bar{\Delta}}
\newcommand{\bx}{\mathbf{x}}
\newcommand{\bz}{\mathbf{z}}
\newcommand{\bw}{\mathbf{w}}
\newcommand{\bbeta}{\boldsymbol{\beta}}
\newcommand{\bSigma}{\boldsymbol{\Sigma}}
\newcommand{\bgamma}{\boldsymbol{\gamma}}

\newcommand{\XTi}[1]{\textcolor{olive}{#1}}

\begin{document}

\title[Quantifying intra-physician variability]{Quantifying intra-physician variability in clinical decision making} 

\author[A. Benani et al.]{
  Alaedine Benani$^{1,2,3,*}$, Pierre Meneton$^{4}$, Emmanuel Messas$^{3}$, \\
  Liza Hettal$^{9,10}$, Sai Sagireddy$^{8}$, Damien Grosgeorge$^{1}$, \\
  Jérôme Salomon$^{1}$, Sylvain Bodard$^{5,6,7}$, Xavier Tannier$^{2}$
}

\thanks{$^*$ Corresponding author: alaedine.benani@aphp.fr}
\date{\today}

\maketitle

\vspace{-0.5cm}
\begin{center}
\scriptsize
$^1$ Preventive Medicine, Data Science and AI Lab, Zoï, F-75010 Paris, France \\
$^2$ Sorbonne Université, Université Sorbonne Paris-Nord, Inserm, Limics, F-75006 Paris, France \\
$^3$ Département cardio-vasculaire, Hôpital européen Georges-Pompidou, université Paris Cité, Inserm UMR 970, F-75015 Paris, France \\
$^4$ Inserm, Sorbonne Université, Université Sorbonne Paris-Nord, Limics, F-75006 Paris, France \\
$^5$ Université de Paris Cité, AP-HP, Hôpital Universitaire Necker Enfants Malades, F-75015 Paris, France \\
$^6$ CNRS UMR 7371, INSERM U 1146, LIB, Sorbonne Université, F-75006 Paris, France \\
$^7$ Massachusetts General Hospital, Harvard Medical School, Boston, MA, USA \\
$^8$ European University Cyprus Frankfurt School of Medicine \\
$^9$ Institut de Cancérologie de Lorraine, Vandoeuvre-lès-Nancy \\
$^{10}$ Université de Lorraine, Nancy, France
\end{center}
\vspace{0.8cm}

\let\thefootnote\relax

\begin{abstract}
Intra-physician prescribing variability, the probability that one physician issues discordant decisions for two patients deemed comparable on observed covariates, holds great impact in quality of care, safety and cost. However, there are no known validated measurement methods. Here, we benchmark eight methods (Euclidean, Mahalanobis, Learned-Weights, Genetic Mahalanobis, Random Forest proximity, Mutual-Information-weighted, Latent Profile Analysis and Bayesian binomial generalized linear mixed model) against a synthetic ground truth across 94 experimental conditions. Learned-Weights matching achieves the lowest mean absolute error ($\overline{|\Dbar|} = 0.027$), followed by Mutual-Information-weighted matching ($0.028$) and RF Proximity ($0.034$). All eight discordance-analysis methods preserve the physician rank ordering with high fidelity (Spearman $\rho \geq 0.89$ versus the ground truth on the SCORE2 experiment), as long as the physician variability groups are well separated. Under a continuous-heterogeneity physician model, rank preservation degrades substantially for unsupervised methods ($\rho \in [0.28, 0.35]$) but is retained by supervised feature-weighted methods and the GLMM ($\rho \in [0.62, 0.68]$). This controlled methodological evaluation is a foundation for validation on observational prescribing data. Once validated on observational prescribing data, these evaluated open-source estimators could turn prescribing inconsistency into a routinely measurable clinician-level quality metric, systematically complementing the existing literature on between-physician variation.
\end{abstract} 

\bigskip


\bigskip

\section{Introduction}
\label{sec:intro}

Variability in prescribing affects healthcare quality, patient safety and cost \citep{Virani2018,Bottle2026,Mousques2010}. Inter-physician variation is extensively studied~\citep{Nyholm2013}; intra-physician variation, defined as the consistency of a single clinician's decisions across comparable patients, receives less methodological attention.

The aim of this paper is to evaluate eight methods to measure the likelihood that a physician issues discordant treatment decisions (e.g., prescribes a statin to one patient but not to another) for two patients deemed clinically comparable on the observed covariates. We name this target intra-physician prescribing inconsistency. Two phenomena are conflated in this definition and cannot be separated from observational data alone: (i) genuine clinical inconsistency, where the physician decides differently on patients identical on all clinically relevant features; and (ii) reliance on features visible to the physician but unobserved by the analyst. We therefore distinguish the entropy of the prescribing process (directly estimable from data) from the intrinsic inconsistency of the physician, a latent reasoning-level attribute that is not identifiable from prescriptions alone and may be due to latent patient heterogeneity or contextual factor. The methods evaluated below are estimators of the former.

To our knowledge, no controlled comparison of analytical methods exists for this problem. Prior work remains limited to imaging contouring \citep{Choi2011,Das2021} and mock decision making \citep{Kobayati2026}.

Our contributions are fourfold.
\begin{enumerate}[label=(\roman*),topsep=0pt,itemsep=2pt]
\item We define an estimand for observable intra-physician prescribing discordance and construct a synthetic data-generating framework in which physician-specific ground-truth discordance is known by construction.

\item We compare eight blind estimation strategies under identical inputs and evaluation rules. Seven are matching-based estimators of discordance rates, whereas the binomial GLMM provides a model-based residual overdispersion score. The methods differ in whether patient comparability is defined from covariates alone or informed by the prescription outcome.

\item We assess calibration and rank preservation in 91~main benchmarking conditions, comprising one SCORE2/SCORE2-OP cardiovascular-risk scenario \citep{score2_2021,score2op_2021} and 90 progressive multi-rule experiments ranging from single-covariate thresholds to high-dimensional conjunctive eligibility rules.

\item We evaluate robustness in three additional SCORE2-based sensitivity analyses, introducing covariate dependence between non-HDL and LDL cholesterol, a non-Gaussian right-skewed HbA1c distribution and a continuous physician heterogeneity cohort, for a total of 94 experimental conditions.
\end{enumerate}

\section{Methods}

\begin{figure*}[htbp]
\centering
\includegraphics[width=\textwidth]{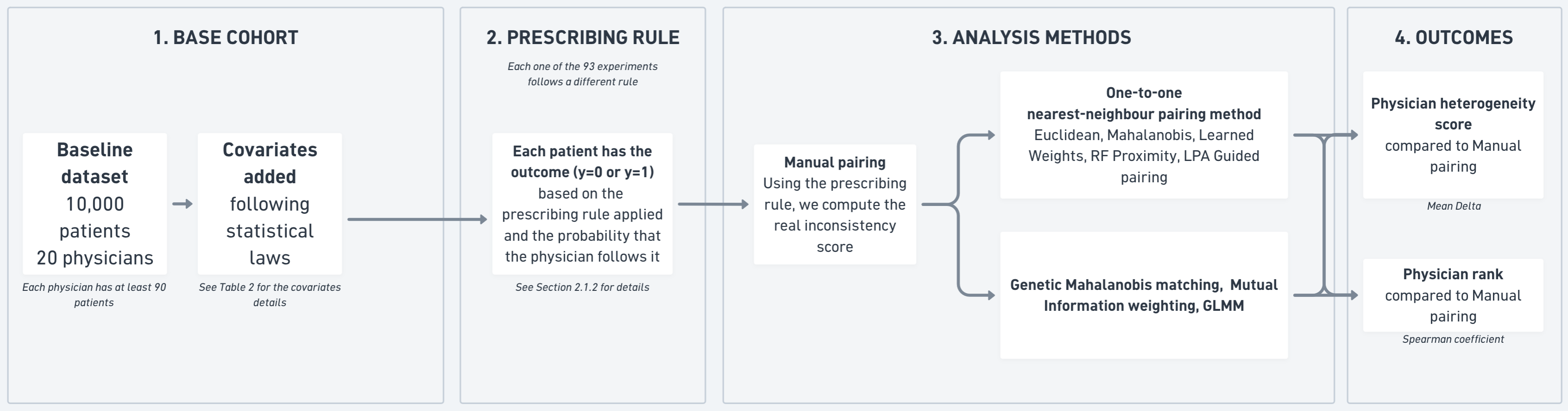}
\caption{\textbf{Overview of the synthetic dataset generation.}}
\label{fig:overview}
\end{figure*}

This section describes the complete simulation and benchmarking workflow used to  evaluate methods for estimating intra-physician prescribing discordance (illustrated by Figure~\ref{fig:overview}). The central difficulty is that, in observational prescribing data, the true set of clinically comparable patients is generally unknown. We therefore use a synthetic framework in which the eligibility rule is known by construction. This allows us to generate prescription outcomes, hide the eligibility rule from the analytical methods, and then compare each method with an oracle reference computed from the known rule.

The workflow proceeds in five steps. First, we generate synthetic cohorts of patients assigned to physicians and described by nine clinical covariates. Second, for each experiment, we define an eligibility rule that determines whether patient $i$ belongs to the target prescribing population, denoted by $m_i = 1$. Eligibility is defined either by a clinically motivated SCORE2/SCORE2-OP cardiovascular-risk rule or by a progressive set of synthetic threshold rules of increasing complexity. Third, we generate the binary prescription outcome $y_i$ from physician-specific Bernoulli probabilities that depend on both the physician and the eligibility status of the patient. This construction induces a known physician-level discordance target.

Fourth, we compute a ground-truth discordance score using the eligibility indicator $m_i$, which is available only because the data are simulated. In parallel, we give the eight discordance-analysis methods only the information that would be available in an observational dataset: the covariates $\bx_i$, the prescription outcome~$y_i$, and the physician identifier $j(i)$. The methods are therefore blind to the eligibility rule. Seven methods estimate a discordance rate by matching patients within physician panels, whereas the GLMM produces a physician-level Pearson-residual overdispersion score.

Fifth, we evaluate method performance along two complementary dimensions. For methods that output a discordance rate, we assess agreement with the ground truth by the mean delta between the estimated and reference discordance scores. For all eight methods, including the GLMM, we assess rank preservation using Spearman's correlation between the method-specific physician scores and the ground-truth physician ranking. 

This study comprises 94 experimental conditions: one SCORE2/SCORE2-OP reference experiment, one separate continuous-heterogeneity SCORE2 benchmark (that evaluates whether the methods still preserve physician rankings when physician behaviour varies along a continuum rather than across well-separated groups), 90 progressive multi-rule experiments, and two additional SCORE2-based sensitivity experiments that assess robustness to covariate dependence and to non-Gaussian covariate distributions. 

The subsections below follow this workflow. Section~\ref{sec:generation} defines the synthetic cohorts, the eligibility rules and the physician prescribing model. Section~\ref{sec:methods} introduces the evaluation metrics and the distinction between covariate-only and outcome-informed methods. Sections~\ref{sec:method_manual} to~\ref{sec:method_glmm} describe the oracle reference, the matching-based methods and the GLMM. Finally, Section~\ref{sec:software} reports implementation and 
reproducibility details.

\subsection{Synthetic data generation}
\label{sec:generation}

Benchmarking intra-physician variability methods requires a known ground truth. We illustrate our approach with a frequent clinical use case: statin prescription in patients at elevated cardiovascular risk. We generate synthetic patient cohorts in which each physician follows a probabilistic prescribing rule conditional on a deterministic eligibility criterion. The ground truth follows from the criterion and the prescribing probabilities. The notation used in this article is defined in Table ~\ref{tab:notation}.

\begin{table*}[htbp]
\centering
\caption{\textbf{Notation used throughout the article}.}
\label{tab:notation}
\footnotesize
\begin{tabular}{@{}cl@{}}
\toprule
Symbol & Meaning \\
\midrule
\multicolumn{2}{@{}l@{}}{\textit{Cohort structure}} \\
$n$                       & Number of patients in the synthetic cohort \\
$J$                       & Number of physicians ($J = 20$ in the main benchmark) \\
$p$                       & Number of covariates ($p = 9$) \\
$i$                       & Patient index, $i = 1, \ldots, n$ \\
$j(i)$                    & Index of the physician seeing patient $i$, $j(i) \in \{1, \ldots, J\}$ \\
$\bx_i \in \mathbb{R}^p$  & Raw covariate vector of patient $i$ \\
$\bz_i$                   & Cohort-wide standardised covariate vector (z-score) \\
$\bz_i^{(R)}$             & Robust-scaled covariate vector (only used for LPA-guided pairing) \\
\midrule
\multicolumn{2}{@{}l@{}}{\textit{Prescribing process}} \\
$y_i \in \{0, 1\}$        & Binary prescription outcome ($1$ if drug prescribed, $0$ otherwise) \\
$m_i \in \{0, 1\}$        & Eligibility flag ($1$ if patient should receive the drug, $0$ otherwise) \\
$p_j^{\text{high}}$       & Probability that physician $j$ prescribes when $m_i = 1$ \\
$p_j^{\text{low}}$        & Probability that physician $j$ prescribes when $m_i = 0$ \\
$D^*_j$                   & Theoretical within-eligible discordance, $D^*_j = 2\,p_j^{\text{high}}(1 - p_j^{\text{high}})$ \\
\midrule
\multicolumn{2}{@{}l@{}}{\textit{Estimation and evaluation}} \\
$\mathcal{M}$              & Set of the eight discordance-analysis methods \\
$\mathcal{M}_{D}$          & Set of discordance-rate methods, excluding the GLMM \\
$S_{j,m}^{(e)}$            & Per-physician score produced by method $m$ in experiment $e$ \\
$\rho_m^{(e)}$             & Spearman rank correlation between method $m$ and the ground truth in experiment $e$ \\
$\mathcal{E}_j$           & Set of eligible patients of physician $j$ \\
$\mathcal{P}_j$           & Set of one-to-one matched patient pairs of physician $j$ \\
$\Dhat_{j,m}$             & Per-physician discordance estimate produced by method $m$ \\
$\Dhat_{j,\text{GT}}$    & Per-physician ground-truth discordance \\
$\Dbar_m^{(e)}$           & Per-experiment mean delta: $J^{-1} \sum_j (\Dhat_{j,m}^{(e)} - \Dhat_{j,\text{GT}}^{(e)})$ \\
$\Dbar_m$                 & Cross-experiment mean delta: $\mathbb{E}_e[\Dbar_m^{(e)}]$ \\
$\overline{|\Dbar_m|}$    & Cross-experiment mean absolute delta \\
$GT$                      & Ground truth \\
\bottomrule
\end{tabular}
\end{table*}

\subsubsection{Base cohort}
\label{sec:gen_base}
The cohort contains $n = 10,000$ patients allocated to $J = 20$ physicians, with at least 90 patients per physician (the remainder is distributed uniformly at random).

Each patient is described by $p = 9$ clinical covariates, listed in  Table~\ref{tab:covariates}. In the main experiments, continuous covariates are sampled independently from Gaussian marginals restricted to clinically plausible ranges, and binary covariates are sampled from Bernoulli distributions. The resulting covariate vector is denoted by $\bx_i \in \mathbb{R}^p$. All covariates are generated before the eligibility rule and prescription outcome are applied.

The main experiments assumes independent covariates and approximately Gaussian continuous marginals. We therefore conduct two additional SCORE2-based sensitivity experiments to assess departures from these assumptions. The first introduced dependence between non-HDL and LDL cholesterol using a Gaussian copula. The second replaced the Gaussian HbA1c marginal by a right-skewed lognormal distribution. These sensitivity experiments are reported in Appendix~\ref{app:sensitivity_data_generation}.

\begin{table*}
\centering
\caption{\textbf{Patient covariates and generating distributions} used in the main 91-experiment benchmark. All covariates are sampled independently. Alternative configurations are reported in Appendix~\ref{app:sensitivity_data_generation}.}
\label{tab:covariates}
\begin{tabular}{@{}llc@{}}
\toprule
Covariate & Distribution & Type \\
\midrule
Age (years)                    & $\mathcal{N}(60, 12^2)$, clipped $[40,90]$        & Integer \\
HbA1c (\%)                     & $\mathcal{N}(6.5, 1.5^2)$, clipped $[4, 12]$      & Continuous \\
Non-HDL cholesterol (mmol/L)   & $\mathcal{N}(3.6, 0.95^2)$, clipped $[1, 8]$      & Continuous \\
HDL cholesterol (mmol/L)       & $\mathcal{N}(1.35, 0.38^2)$, clipped $[0.4, 2.8]$ & Continuous \\
LDL cholesterol (g/L)          & $\mathcal{N}(1.3, 0.35^2)$, clipped $[0.4, 2.2]$  & Continuous \\
Systolic BP (mmHg)             & $\mathcal{N}(130, 20^2)$, clipped $[90, 200]$     & Integer \\
eGFR (mL/min/1.73\,m$^2$)      & $\mathcal{N}(90, 25^2)$, clipped $[15, 140]$      & Continuous \\
Smoker                         & $\text{Bernoulli}(0.20)$                          & Binary \\
Male sex                       & $\text{Bernoulli}(0.60)$                          & Binary \\
\bottomrule
\end{tabular}
\end{table*}

\subsubsection{Prescribing rule and physician model}
\label{sec:gen_prescribing}

Each patient\XTi{~$i$} receives a prescription eligibility flag $m_i \in \{0,1\}$. The prescription outcome $y_i \in \{0,1\}$ is then drawn from a physician- and eligibility-specific Bernoulli:
\begin{equation}
\label{eq:outcome}
y_i \mid j(i), m_i \sim
\begin{cases}
\text{Bern}(p_j^{\text{high}}) & \text{if } m_i = 1, \\
\text{Bern}(p_j^{\text{low}})  & \text{if } m_i = 0.
\end{cases}
\end{equation}
The 20 physicians form five groups of four, each defined by a fixed pair $(p^{\text{high}}, p^{\text{low}})$ (Table~\ref{tab:physician_groups}). A random pair of eligible patients in group $g$ is discordant, meaning that one patient has a prescription but the other patient does not have a prescription, with probability $D^* = 2p_g^{\text{high}}(1 - p_g^{\text{high}})$.

\begin{table}
\centering
\caption{Physician behaviour groups and expected ground-truth discordance $D^* = 2p^{\text{high}}(1-p^{\text{high}})$.}
\label{tab:physician_groups}
\begin{tabular}{@{}lcccc@{}}
\toprule
Group & Phys. & $p^{\text{high}}$ & $p^{\text{low}}$ & $D^*$ \\
\midrule
1 (Deterministic)   & P1--P4   & 1.00 & 0.00 & 0.000 \\
2 (Near-determ.)    & P5--P8   & 0.90 & 0.05 & 0.180 \\
3 (Moderate)        & P9--P12  & 0.80 & 0.10 & 0.320 \\
4 (Substantial)     & P13--P16 & 0.70 & 0.20 & 0.420 \\
5 (Coin-flip)       & P17--P20 & 0.50 & 0.50 & 0.500 \\
\bottomrule
\end{tabular}
\end{table}

\textbf{Quantile-calibrated thresholds.} We use a quantile-calibrated rule: each experiment draws $p^\star \sim \mathcal{U}(0.2, 0.8)$ independently and sets the per-covariate threshold to
\begin{equation}
\tau_\ell = Q_{f_\ell}\bigl((p^\star)^{1/w}\bigr).
\end{equation}
Under independence of the covariates, $\Pr(m_i=1) \approx p^\star$ regardless of $w$, and every one of the 94 experiments has a non-empty eligible stratum within every physician panel. The independent draws of $p^\star$ and of the per-covariate thresholds make the two passes statistically independent realisations at every window position.

\paragraph{SCORE2 reference experiment}
The reference experiment uses SCORE2/SCORE2-OP \citep{score2_2021,score2op_2021}, a cardiovascular risk score routinely used in clinical practice. A patient is eligible when the 10-year cardiovascular risk meets the moderate-to-high risk threshold:
\begin{equation}
m_i =
\begin{cases}
1 & \text{age} < 50,\; \text{risk} \geq 2.5\%, \\
1 & 50 \leq \text{age} \leq 69,\; \text{risk} \geq 5\%, \\
1 & \text{age} > 69,\; \text{risk} \geq 7.5\%, \\
0 & \text{otherwise}.
\end{cases}
\end{equation}
SCORE2/SCORE2-OP is computed from age, sex, smoking status, systolic blood pressure, total cholesterol (computed as non-HDL $+$ HDL) and HDL cholesterol with the published ``Low'' risk-region coefficients \citep{score2op_2021}. For convenience, the experiment using SCORE2/SCORE2-OP is referred to as the "SCORE2 experiment".

SCORE2 therefore uses six of the nine covariates (age, sex, smoking, SBP, total cholesterol $=$ non-HDL $+$ HDL, HDL); the other three (HbA1c, LDL, eGFR) are noise.

\paragraph{Continuous-heterogeneity SCORE2 experiment}
\label{sec:methods_continuous}

The main SCORE2 experiment uses a five-group physician model in which physicians occupy five well-separated behavioural levels of $p^{\text{high}}$ (Table~\ref{tab:physician_groups}). This design provides a calibration benchmark: it tests whether the methods recover known discordance levels when between-physician differences are deliberately separable. To assess the more deployment-relevant discrimination problem, we also define a continuous-heterogeneity SCORE2 benchmark, in which physicians vary along a continuum rather than across discrete behavioural groups.

The continuous-heterogeneity cohort contains $n = 20\,000$ patients distributed across $J = 50$ physicians. For each physician~$j$, the high-eligibility prescription probability is drawn independently as
\[
p_j^{\text{high}} \sim \mathcal{U}(0.5, 1.0),
\]
whereas $p_j^{\text{low}}$ is fixed at $0.05$ for every physician. The eligibility rule is the same SCORE2 / SCORE2-OP rule as in the reference SCORE2 experiment. Prescription outcomes are generated according to Equation~\ref{eq:outcome}.

For physician $j$, the theoretical within-eligible discordance is
\begin{equation}
D^\star_j = 2p_j^{\text{high}}(1 - p_j^{\text{high}}).
\end{equation}

All analytical methods use the same implementation and hyperparameters as in the main benchmark. Genetic Mahalanobis is omitted from this experiment because of its computational cost, approximately $10$~min per experiment under the current implementation.

This benchmark is analysed separately from the 93-condition cross-experiment benchmark because it changes both the physician-generating model and the physician panel size. Its primary endpoint is rank discrimination. For each method $m$, we compute Spearman's rank correlation between the per-physician method score $S_{j,m}$ and the true discordance $D^\star_j$ across the $J=50$ physicians. For discordance-rate methods, $S_{j,m} = \Dhat_{j,m}$; for the GLMM, $S_{j,m} = \widehat{\mathrm{OD}}_j$.

\paragraph{Progressive multi-rule experiments}
\label{sec:gen_progressive}

These experiences are designed to evaluate the robustness of discordance‑estimation methods under alternative definitions of the eligible patient population. By progressively increasing the number of covariates that determine eligibility (a multi‑rule, progressive design) and by introducing different dependency structures, we examine whether method performance is sensitive to the size of the eligibility region, the degree of heterogeneity, and potential distributional misspecification.

Let $\mathcal{F} = (f_1, \ldots, f_9)$ be the ordered covariate list. For each window size $w \in \{1, \ldots, 9\}$ a window of $w$ consecutive covariates slides across $\mathcal{F}$, producing $9 - w + 1$ window positions. Each window position defines one experiment whose eligibility rule is a conjunction:
\begin{equation}
m_i = \prod_{\ell \in W} \mathbb{1}[f_\ell(i) \leq \tau_\ell],
\end{equation}
with $W$ the active window and $\tau_\ell$ a quantile-calibrated threshold.

\paragraph{Two-pass design}
A full sweep over all window sizes and positions yields $\sum_{w=1}^{9} (9 - w + 1) = 45$ rules. We replicate this sweep over 2 independent passes. For each experiment, the two passes share the rule structure (same active covariates, same $w$) but use independent random seeds for both the per-covariate thresholds and the base cohort. The two passes therefore differ in (i)~the realised thresholds $\tau_\ell$ and (ii)~the patient draws, while keeping the generating distributions (Table~\ref{tab:covariates}) and the physician behaviour model (Table~\ref{tab:physician_groups}) fixed. Comparing pass~1 to pass~2 isolates the contribution of joint threshold-plus-cohort sampling variability to the cross-experiment results.
The design covers single-covariate thresholds ($w=1$) that probe robustness to irrelevant covariates, moderate-complexity conjunctive rules (\mbox{$w=2$--$4$}), and high-dimensional regions ($w=5$--$9$).

\subsection{Analytical methods}
\label{sec:methods}

\subsubsection{Overview}
\label{sec:methods_overview}

We evaluate eight methods. The ground truth is elaborated by eligibity-rule manual-pairing. All methods but the Bayesian binomial GLMM are matching-based: each constructs a per-physician distance matrix and forms patient pairs using one-to-one nearest-neighbour matching under a caliper (Algorithm~\ref{alg:hungarian_caliper}). The genetic matching is the only method that uses nearest-neighbour matching with replacement. All take the same input: $n$ patients with \mbox{$p = 9$} covariates $\bx_i \in \mathbb{R}^p$, a binary outcome $y_i$, and a physician identifier $j(i)$. Each produces a per-physician variability score.

The eight discordance-analysis methods are blind to the eligibility rule. 

\subsubsection{Delta to the ground truth}
\label{sec:methods_ground_truth}

First, for the seven methods that output a discordance rate, we quantify absolute agreement using the mean delta against the ground truth. The GLMM is evaluated separately because its Pearson-residual overdispersion score is not on the discordance-rate scale.

Let $\mathcal{M}_{D}$ denote the set of discordance-rate methods, indexed by $m$. 
For experiment $e$, physician $j \in \{1,\ldots,J\}$, and method $m \in \mathcal{M}_{D}$, let $\Dhat_{j,m}^{(e)}$ be the discordance rate estimated by method $m$ for physician $j$. 
Let $\Dhat_{j,\mathrm{GT}}^{(e)}$ be the corresponding ground-truth discordance, computed from the known eligibility rule. 
The per-experiment mean delta for method $m$ is
\begin{equation}
\label{eq:delta}
\Dbar_m^{(e)} 
= \frac{1}{J} \sum_{j=1}^{J} 
\left(
\Dhat_{j,m}^{(e)} - \Dhat_{j,\mathrm{GT}}^{(e)}
\right),
\qquad J = 20.
\end{equation}
Across experiments, the aggregate mean delta is the empirical average
\begin{equation}
\Dbar_a 
= \frac{1}{|\mathcal{E}|} \sum_{e \in \mathcal{E}} \Dbar_m^{(e)},
\end{equation}
where $\mathcal{E}$ is the set of benchmark experiments. 
Positive values indicate average overestimation of physician inconsistency relative to the ground truth, negative values indicate average underestimation, and zero indicates no average bias across physicians and experiments.

\subsubsection{Rank preservation}
\label{sec:methods_rank}

Second, for all eight discordance-analysis methods, including the GLMM, we assess rank preservation using Spearman's rank correlation with the ground truth. 
Let $\mathcal{A}$ denote the full set of eight methods. 
For method $a \in \mathcal{A}$, define the per-physician method score
\[
S_{j,m}^{(e)} =
\begin{cases}
\Dhat_{j,m}^{(e)} & \text{for discordance-rate methods}, \\
\widehat{\mathrm{OD}}_{j}^{(e)} & \text{for the GLMM}.
\end{cases}
\]
The rank-preservation coefficient is
\begin{equation}
\label{eq:spearman}
\rho_a^{(e)}
=
\operatorname{corr}_{\mathrm{Spearman}}
\left(
\left(S_{1,m}^{(e)},\ldots,S_{J,m}^{(e)}\right),
\left(\Dhat_{1,\mathrm{GT}}^{(e)},\ldots,\Dhat_{J,\mathrm{GT}}^{(e)}\right)
\right).
\end{equation}
Spearman's $\rho$ is scale-free and therefore allows comparison between discordance-rate methods and the GLMM Pearson-residual overdispersion score. 
Values close to $1$ indicate that a method preserves the physician ordering induced by the ground truth.

\subsubsection{Confidence intervals}
Unless stated otherwise, confidence intervals reported in this article are $95\%$ percentile bootstrap intervals with $B = 2000$ resamples and seed $42$. The resampling unit is the physician for per-physician statistics (Table~\ref{tab:score2_groups}) and the experiment for cross-experiment statistics (Tables~\ref{tab:cross_experiment}--\ref{tab:passes}).

\subsubsection{Unsupervised versus supervised methods}
\label{sec:methods_supervision}
Three methods are unsupervised (use $\bx_i$ only): Euclidean, Mahalanobis, LPA-guided. Five are supervised (use both $\bx_i$ and $y_i$): Learned Weights, Genetic Mahalanobis, RF Proximity, Mutual Information, and GLMM. ``Blind'' refers only to the eligibility rule $m_i$, which is given solely to the manual pairing for ground truth set up. Supervised methods may exhibit a feedback loop: by learning weights or similarities from $y$ they pull patients with similar $y$ closer in the comparability space, reducing the matched discordance rate.

\subsection{Manual Pairing for the ground truth}
\label{sec:method_manual}
For each physician $j$, let $\mathcal{E}_j = \{i : j(i)=j,\; m_i=1\}$. The ground-truth discordance rate is
\begin{equation}
\Dhat_{j,\text{GT}} = \frac{\sum_{i<k,\, i,k \in \mathcal{E}_j} \mathbb{1}[y_i \neq y_k]}{\binom{|\mathcal{E}_j|}{2}},
\end{equation}
computed over all pairs of eligible patients. In the synthetic framework, $\Dhat_{j,\text{GT}}$ converges to $2 p_j^{\text{high}}(1-p_j^{\text{high}})$ and coincides with the physician's intrinsic inconsistency by construction. In observational data, the analogous quantity estimates the entropy of the prescribing process, not necessarily reasoning-level inconsistency.

\subsection{One-to-one nearest-neighbour pairing methods}
\label{sec:method_nn}
Methods share both their distance construction (on z-scored covariates) and their pairing procedure. From a per-physician distance matrix ${D}_j$, one-to-one pairs are formed by Hungarian assignment followed by undirected-edge deduplication and weight-sorted greedy selection without replacement, under a Rosenbaum Rubin caliper at the 25th percentile of within-physician pairwise distances \citep{rosenbaum1985caliper} (Algorithm~\ref{alg:hungarian_caliper}). The Hungarian-then-greedy procedure is a heuristic for the minimum-weight one-to-one matching problem. A pure greedy nearest-neighbour fallback is retained in the code base but is not used in the runs reported here. The same pairing procedure and caliper are used by LPA-guided pairing and Mutual Information weighting; Genetic Mahalanobis uses a different pairing rule, described in Section~\ref{sec:method_genmah}.

\begin{algorithm*}[htbp]
\caption{\textbf{One-to-one nearest-neighbour pairing with caliper} (one physician). Hungarian assignment followed by deduplicated greedy edge selection.}
\label{alg:hungarian_caliper}
\begin{algorithmic}[1]
\State Set $\mathbf{D}_j' \gets \mathbf{D}_j$ with $d_{ii}' \gets +\infty$ on the diagonal.
\State Compute $c_j \gets$ 25th percentile of the upper-triangular off-diagonal entries of $\mathbf{D}_j$.
\State Solve the Hungarian assignment problem on $\mathbf{D}_j'$:
\Statex \hspace{1em} $(\sigma_1, \ldots, \sigma_{n_j}) \gets \texttt{linear\_sum\_assignment}(\mathbf{D}_j')$.
\State Build the undirected candidate set $\mathcal{C} \gets \emptyset$.
\For{$i = 1, \ldots, n_j$}
    \State Let $(a, b) \gets (\min(i, \sigma_i), \max(i, \sigma_i))$ and $w \gets d_{ab}$.
    \If{$a \neq b$ \textbf{and} $w$ is finite \textbf{and} $w \leq c_j$}
        \State $\mathcal{C} \gets \mathcal{C} \cup \{(w, (a,b))\}$.
    \EndIf
\EndFor
\State Sort $\mathcal{C}$ by ascending $w$ and deduplicate identical $(a,b)$.
\State Initialise $\mathcal{P} \gets \emptyset$; $\mathcal{U} \gets \emptyset$ (used patients).
\For{each $(w, (a, b)) \in \mathcal{C}$ in ascending order}
    \If{$a \in \mathcal{U}$ \textbf{or} $b \in \mathcal{U}$} \textbf{continue} \EndIf
    \State $\mathcal{P} \gets \mathcal{P} \cup \{(a, b)\}$;\ $\mathcal{U} \gets \mathcal{U} \cup \{a, b\}$.
\EndFor
\State \Return $\mathcal{P}$
\end{algorithmic}
\end{algorithm*}

The discordance rate is $\Dhat_{j,m} = |\mathcal{P}_j|^{-1} \sum_{(i,k) \in \mathcal{P}_j} \mathbb{1}[y_i \neq y_k]$. Covariates are standardised cohort-wide,
\begin{equation}
\bz_i = (\bx_i - \bar{\bx})/\boldsymbol{\sigma},
\end{equation}
with $\bar{\bx}$ and $\boldsymbol{\sigma}$ the cohort-wide mean and standard-deviation vectors.

\subsubsection{Euclidean}
\label{sec:method_eucl}
$d_{ik}^{\text{Eucl}} = \|\bz_i - \bz_k\|_2 / \sqrt{p}$.

\subsubsection{Mahalanobis}
\label{sec:method_mah}
$d_{ik}^{\text{Mah}} = \sqrt{(\bz_i - \bz_k)^\top \hat{\bSigma}^{-1}(\bz_i - \bz_k)}$, with $\hat{\bSigma}$ the standardised-covariate covariance, inverted by Moore--Penrose pseudo-inverse.

\subsubsection{Learned Weights}
\label{sec:method_learn}
A random forest ($T=300$ trees, max depth 8, $\text{min\_samples\_leaf} = \max(5, \lfloor 0.01\,n \rfloor)$) is trained to predict $y$ from $z$. Its normalised Gini importances $\bgamma$ rescale the standardised space:
\begin{equation}
d_{ik}^{\text{Learn}} = \bigl\|\text{diag}(\sqrt{\bgamma}) (\bz_i - \bz_k)\bigr\|_2.
\end{equation}

\subsubsection{RF Proximity}
\label{sec:method_rf}
A RF ($T=300$, max depth 8, $\text{min\_samples\_leaf} = \max(5, \lfloor 0.01\,n \rfloor)$) is trained to predict $y$ from $\bx$. The proximity is the fraction of trees in which two patients share a terminal node:
\begin{equation}
\text{prox}(i,k) = T^{-1} \sum_t \mathbb{1}[\ell_t(i) = \ell_t(k)],
\end{equation}
and the dissimilarity is $d_{ik}^{\text{RF}} = 1 - \text{prox}(i,k)$ \citep{breiman2001random,shi2006unsupervised}.

\subsection{Genetic Mahalanobis matching method}
\label{sec:method_genmah}
This method optimises covariate weights by genetic search and applies them inside a Mahalanobis distance \citep{diamond2013genetic}. A purely Euclidean genetic variant is discarded because it yields near-identical results.

\subsubsection{Weight optimisation}
Differential evolution searches $\bw \in [0, 10]^p$ for the minimiser of
\begin{equation}
\mathcal{L}(\bw) = \max_\ell \bigl|\bar z_\ell^{(1)} - \bar z_\ell^{(0,\text{matched})}\bigr|,
\end{equation}
the maximum absolute mean difference, in the standardized covariates space, between treated ($y=1$) patients and their matched controls. For each candidate $\bw$ the objective normalizes $\tilde \bw = \bw / |\bw|_1$, builds the weighted space $\tilde \bz_i = \text{diag}(\sqrt{\tilde \bw}),\bz_i$, and matches each treated patient to its single nearest control by Euclidean distance, with replacement. Euclidean rather than Mahalanobis distance is used inside the optimization loop for tractability: differential evolution evaluates $\sim 1,500$ candidate weight vectors, and a Mahalanobis matching would require re-inverting the weighted-space covariance at every evaluation. The Mahalanobis distance is reserved for the per-physician matching at Stage 2 below, where the inversion is performed only once per physician.

\subsubsection{Application}
The optimized weights $\hat \bw$ are now applied inside a Mahalanobis distance, which accounts for residual covariates correlation in the per-physician matching. With $\tilde \bz_i = \text{diag}(\sqrt{\hat \bw}) \bz_i$, the per-physician distance is
\begin{equation}
d_{ik}^{\text{GM}} = \sqrt{(\tilde \bz_i - \tilde \bz_k)^\top \hat{\bSigma}_{\tilde z}^{-1}(\tilde \bz_i - \tilde \bz_k)}.
\end{equation}

\subsubsection{Pairing}
Genetic Mahalanobis does not apply Algorithm~\ref{alg:hungarian_caliper}. Each patient $i$ is paired with its single nearest neighbour $\mathrm{nn}_j(i) = \arg\min_{k \neq i} d_{ik}^{\text{GM}}$ inside the physician panel; the same neighbour may serve multiple patients (matching with replacement). The Rosenbaum--Rubin caliper at the 25th percentile is applied: patient $i$ is kept only if $d_{i,\mathrm{nn}_j(i)}^{\text{GM}} \leq c_j$. The discordance rate is
\begin{equation}
\Dhat_{j,\text{GM}} = \frac{1}{|\mathcal{K}_j|} \sum_{i \in \mathcal{K}_j} \mathbb{1}[y_i \neq y_{\mathrm{nn}_j(i)}],
\end{equation}
where $\mathcal{K}_j$ is the set of patients whose nearest neighbour passes the caliper.

\subsection{Latent Profile Analysis-guided pairing method}
\label{sec:method_lpa}
LPA-guided pairing matches patients in the membership space of a Gaussian mixture fitted globally on the covariate matrix \citep{vermunt2002latent}.

\subsubsection{Pre-processing}
The cohort-wide standardisation of Section~\ref{sec:method_nn} is replaced by a robust scaler (median-centred, IQR-scaled; z-score fallback for columns with zero post-scaling variance). We denote the robust-scaled covariate vector $\bz_i^{(R)}$. The Gaussian mixture and the LPA distance below both operate on $\bz_i^{(R)}$.

\subsubsection{Latent profile selection}
For each $K \in \{2, \ldots, 10\}$, a full-covariance Gaussian mixture is fitted to $\{\bz_i^{(R)}\}$ with $n_\text{init} = 10$ initialisations and covariance regularisation $10^{-6}$. The model with the lowest BIC among the converged candidates is retained, $\hat{\mathcal{M}}_K$; $K=2$ is the fallback when no candidate converges. Each patient is described by a posterior membership $\boldsymbol{\pi}_i \in \Delta^{K-1}$.

\subsubsection{Hybrid latent--clinical distance}
Per physician, the latent and clinical pairwise distances are
\begin{equation}
d_{ik}^{\text{lat}} = \|\boldsymbol{\pi}_i - \boldsymbol{\pi}_k\|_2,\qquad
d_{ik}^{\text{cli}} = \|\bz_i^{(R)} - \bz_k^{(R)}\|_2.
\end{equation}
Each is rescaled by its physician-level maximum and combined:
\begin{equation}
d_{ik}^{\text{LPA}} = \alpha\,\frac{d_{ik}^{\text{lat}}}{\max d^{\text{lat}}} + (1-\alpha)\,\frac{d_{ik}^{\text{cli}}}{\max d^{\text{cli}}},\qquad \alpha = 0.5.
\end{equation}
The clinical term breaks ties when several patients share nearly identical memberships. The one-to-one pairing of Algorithm~\ref{alg:hungarian_caliper} (Hungarian solver) and the caliper of Section~\ref{sec:method_nn} are applied to $\mathbf{D}_j^{\text{LPA}}$.

\subsection{Mutual Information weighting}
\label{sec:method_mi}
Weights $\tilde w_\ell = I(z_\ell; y)/\sum_{\ell'} I(z_{\ell'}; y)$ are estimated by $k$-nearest-neighbour mutual information \citep{kraskov2004estimating}. The distance is
\begin{equation}
d_{ik}^{\text{MI}} = \bigl\|\text{diag}(\sqrt{\tilde\bw}) (\bz_i - \bz_k)\bigr\|_2,
\end{equation}
followed by the one-to-one pairing of Algorithm~\ref{alg:hungarian_caliper} and the caliper of Section~\ref{sec:method_nn}.

\subsection{Bayesian binomial GLMM with Pearson-residual overdispersion}
\label{sec:method_glmm}
Unlike the matching-based methods, the GLMM does not rely on pairing. It is a supervised method: its parameters are fitted to $y$.

\subsubsection{Model}
With $j(i)$ the physician of patient $i$ and an intercept $\beta_0$,
\begin{equation}
\text{logit}\,\Pr(y_i = 1 \mid \bz_i, u_{j(i)}) = \beta_0 + \bbeta^\top \bz_i + u_{j(i)}, \qquad u_j \overset{\text{iid}}{\sim} \mathcal{N}(0, \sigma_u^2).
\end{equation}
The random intercept $u_j$ captures inter-physician variation. Estimation uses mean-field variational Bayes \citep{blei2017variational,seabold2010statsmodels}.

\subsubsection{Per-physician variability score}
For each patient, $\hat p_i = \mathrm{logit}^{-1}(\hat\beta_0 + \hat\bbeta^\top \bz_i + \hat u_{j(i)})$ and
\begin{equation}
r_i = \frac{y_i - \hat p_i}{\sqrt{\hat p_i(1 - \hat p_i)}}, \qquad
\widehat{\mathrm{OD}}_j = \frac{1}{n_j} \sum_{i:\,j(i)=j} r_i^2.
\end{equation}
A high $\widehat{\mathrm{OD}}_j$ flags physicians whose decisions remain poorly explained by the GLMM after adjustment for covariates and global tendency.

\subsubsection{Bernoulli caveat}
The Bernoulli mean--variance is constrained: $\mathrm{Var}(y_i) = p_i(1-p_i)$, with no free dispersion parameter. $\widehat{\mathrm{OD}}_j$ is therefore a heuristic ranking score, not a calibrated variance. Identifiable Bernoulli overdispersion requires further structural assumptions (an informative prior on an observation-level random effect; a hierarchical Bernoulli--Beta) that we do not pursue.

\subsubsection{Conceptual distinction}
Matching-based methods estimate the empirical probability that two comparable patients receive discordant decisions. The GLMM quantifies unexplained residual variation under a logistic-additive regression. Convergent rankings may indicate that both signals track a common inconsistency, but the constructs are not equivalent. Pearson-residual diagnostics and the calibration curve are in Appendices~\ref{app:pearson_residuals} and~\ref{app:calibration_curve}.

\subsection{Software and reproducibility}
\label{sec:software}
The pipeline is implemented in Python 3.11 (\texttt{numpy}, \texttt{scipy}, \texttt{scikit-learn}, \texttt{statsmodels}). LPA uses \texttt{sklearn.mixture.GaussianMixture}. The GLMM is fitted with \texttt{genmod.bayes\_mixed\_glm.BinomialBayesMixedGLM.fit\_vb}. Genetic matching uses \texttt{optimize.differential\_evolution} in parallel. Random number generators are seeded deterministically. Unit tests cover synthetic data generation, distance and matching procedures, GLMM fitting and LPA selection. Code and configurations are available on GitHub\footnote{\url{https://github.com/alaedinebe/physician-intra-variability}}.

\section{Results}
\label{sec:results}

The manual pairing computes the ground truth: the five physician groups recover their theoretical discordances $2p(1-p) \in \{0, 0.18, 0.32, 0.42, 0.50\}$ to within Monte-Carlo noise: 0.000, 0.174, 0.295, 0.420, 0.502.

\subsection{SCORE2 reference experiment}
\label{sec:results_score2}
\begin{figure*}[htbp]
\centering
\includegraphics[width=\textwidth]{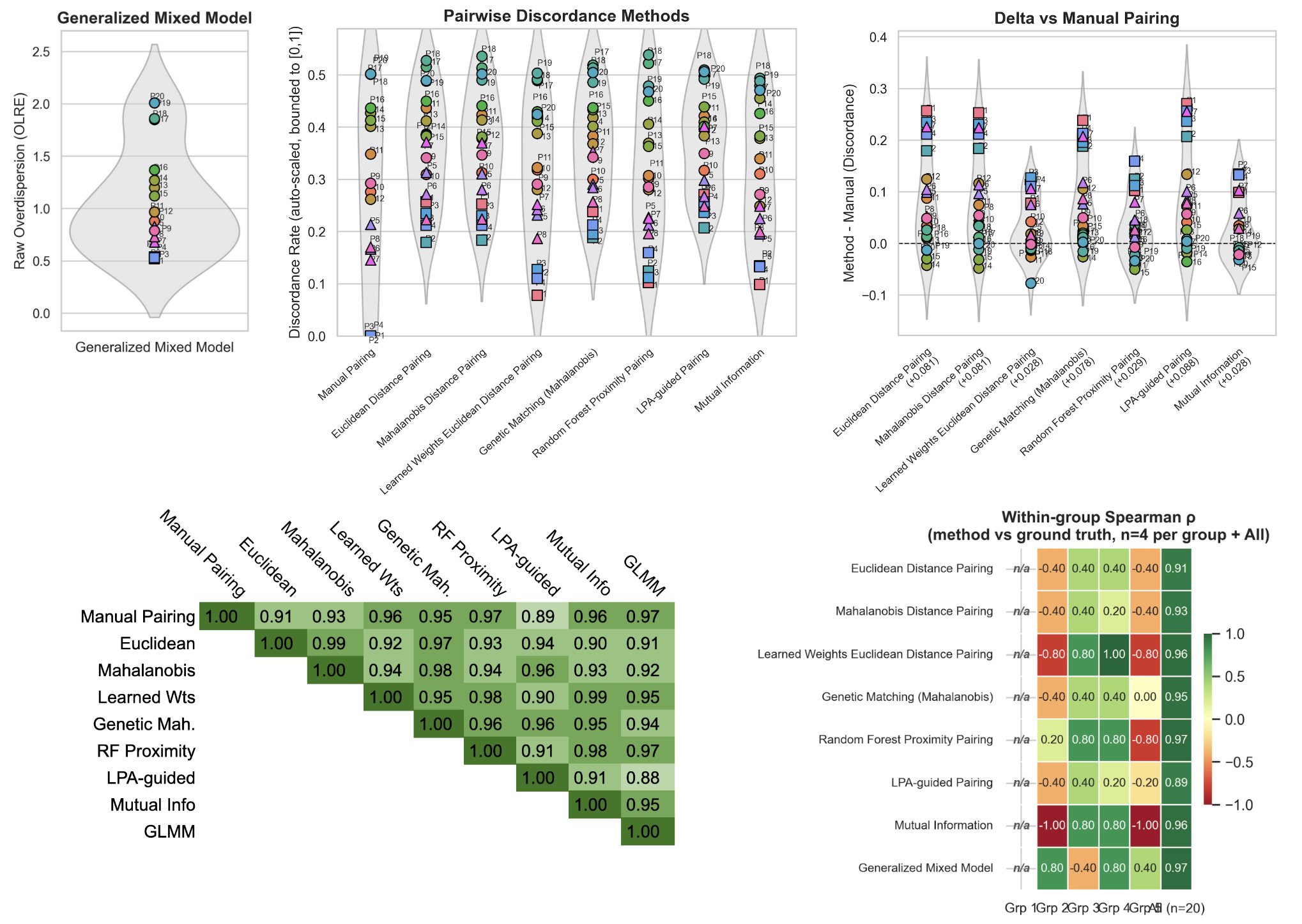}
\caption{\textbf{Physician inconsistency across all methods on the SCORE2 experiment ($J=20$).} \\ \textit{Top-Left:} per-physician GLMM Pearson-residual overdispersion $\widehat{\mathrm{OD}}_j$. \\ \textit{Top-Centre:} per-method discordance rates, coloured by physician group ($p^{\text{high}} \in \{1.00, 0.90, 0.80, 0.70, 0.50\}$). \\ \textit{Top-right:} per-physician $\Delta = \text{method} - \text{ground truth}$; dashed line at zero indicates perfect agreement; values in parentheses are the mean $\Dbar$. \\ \textit{Bottom:} Rank-correlation matrix.}
\label{fig:score2_results}
\end{figure*}

Figure~\ref{fig:score2_results} reports the SCORE2 experiment. 

All matching methods overestimate discordance ($\Dbar > 0$), most strongly in low-heterogeneity groups (Figure~\ref{fig:score2_results}, Table~\ref{tab:score2_groups}). Feature-weighted methods (Learned Weights $+0.028$; RF Proximity $+0.029$; Mutual Information $+0.028$) outperform unweighted ones (Euclidean $+0.081$; Mahalanobis $+0.081$; LPA-guided $+0.088$; Genetic Mahalanobis $+0.078$).

\begin{table*}[htbp]
\centering
\caption{\textbf{Mean discordance rate by physician group and method on the SCORE2 experiment}, with $95\%$ percentile bootstrap confidence intervals ($B = 2000$ resamples, seed 42). Per-group cells resample the four physicians of each group; the $\Dbar$ column resamples all $J = 20$ physicians. Per-group CIs are wide because each group contains only four physicians. The GLMM is not diplayed here because its Pearson-residual score is not on the discordance-rate scale.}
\label{tab:score2_groups}
\scriptsize
\setlength{\tabcolsep}{3pt}
\resizebox{\textwidth}{!}{%
\begin{tabular}{@{}lcccccc@{}}
\toprule
Approach & Grp\,1 & Grp\,2 & Grp\,3 & Grp\,4 & Grp\,5 & $\Dbar$ \\
\midrule
Ground truth
 & 0.000 [0.00,\,0.00]
 & 0.174 [0.15,\,0.20]
 & 0.295 [0.27,\,0.33]
 & 0.420 [0.41,\,0.43]
 & 0.502 [0.50,\,0.50]
 & --- \\
Euclidean
 & 0.221 [0.19,\,0.25]
 & 0.295 [0.25,\,0.35]
 & 0.369 [0.33,\,0.41]
 & 0.408 [0.38,\,0.43]
 & 0.506 [0.49,\,0.52]
 & $+$0.081 [$+$0.04,\,$+$0.12] \\
Mahalanobis
 & 0.220 [0.20,\,0.24]
 & 0.296 [0.25,\,0.35]
 & 0.366 [0.33,\,0.40]
 & 0.404 [0.38,\,0.43]
 & 0.511 [0.50,\,0.53]
 & $+$0.081 [$+$0.04,\,$+$0.12] \\
Learned Wts
 & 0.107 [0.09,\,0.12]
 & 0.228 [0.20,\,0.25]
 & 0.303 [0.29,\,0.32]
 & 0.412 [0.40,\,0.43]
 & 0.478 [0.44,\,0.50]
 & $+$0.028 [$+$0.01,\,$+$0.05] \\
Genetic Mah.
 & 0.209 [0.19,\,0.23]
 & 0.297 [0.27,\,0.34]
 & 0.349 [0.32,\,0.38]
 & 0.423 [0.41,\,0.44]
 & 0.506 [0.49,\,0.52]
 & $+$0.078 [$+$0.04,\,$+$0.12] \\
RF Proximity
 & 0.125 [0.11,\,0.15]
 & 0.216 [0.20,\,0.23]
 & 0.294 [0.28,\,0.31]
 & 0.397 [0.37,\,0.43]
 & 0.502 [0.47,\,0.53]
 & $+$0.029 [$+$0.01,\,$+$0.05] \\
LPA-guided
 & 0.243 [0.22,\,0.26]
 & 0.304 [0.26,\,0.37]
 & 0.371 [0.33,\,0.41]
 & 0.409 [0.39,\,0.43]
 & 0.502 [0.50,\,0.51]
 & $+$0.088 [$+$0.05,\,$+$0.13] \\
Mutual Info
 & 0.125 [0.11,\,0.14]
 & 0.217 [0.20,\,0.24]
 & 0.293 [0.26,\,0.33]
 & 0.411 [0.38,\,0.44]
 & 0.483 [0.48,\,0.49]
 & $+$0.028 [$+$0.00,\,$+$0.05] \\
\bottomrule
\end{tabular}
}%
\end{table*}

\textbf{Rank preservation.} Spearman correlations with the ground truth range from $\rho=0.89$ (LPA-guided) to $\rho=0.97$ (RF Proximity) (Fig~\ref{fig:score2_results}). The supervised feature-weighted methods, meaning RF Proximity ($\rho=0.97$), GLMM ($\rho=0.97$), Learned Weights ($\rho=0.96$) and Mutual Information ($\rho=0.96$), lead. The unsupervised methods trail by a small but visible margin.

\subsection{Continuous-heterogeneity SCORE2 experiment}
\label{sec:results_continuous}

The 5-group physician model used throughout the main experiment (Table~\ref{tab:physician_groups}) places physicians on five well-separated rungs of $p^{\text{high}}$. To probe how the methods behave when physicians lie on a continuum rather than in discrete bins, we replicate the SCORE2 experiment with a continuous-heterogeneity physician model.

\begin{figure*}[htbp]
\centering
\includegraphics[width=\textwidth]{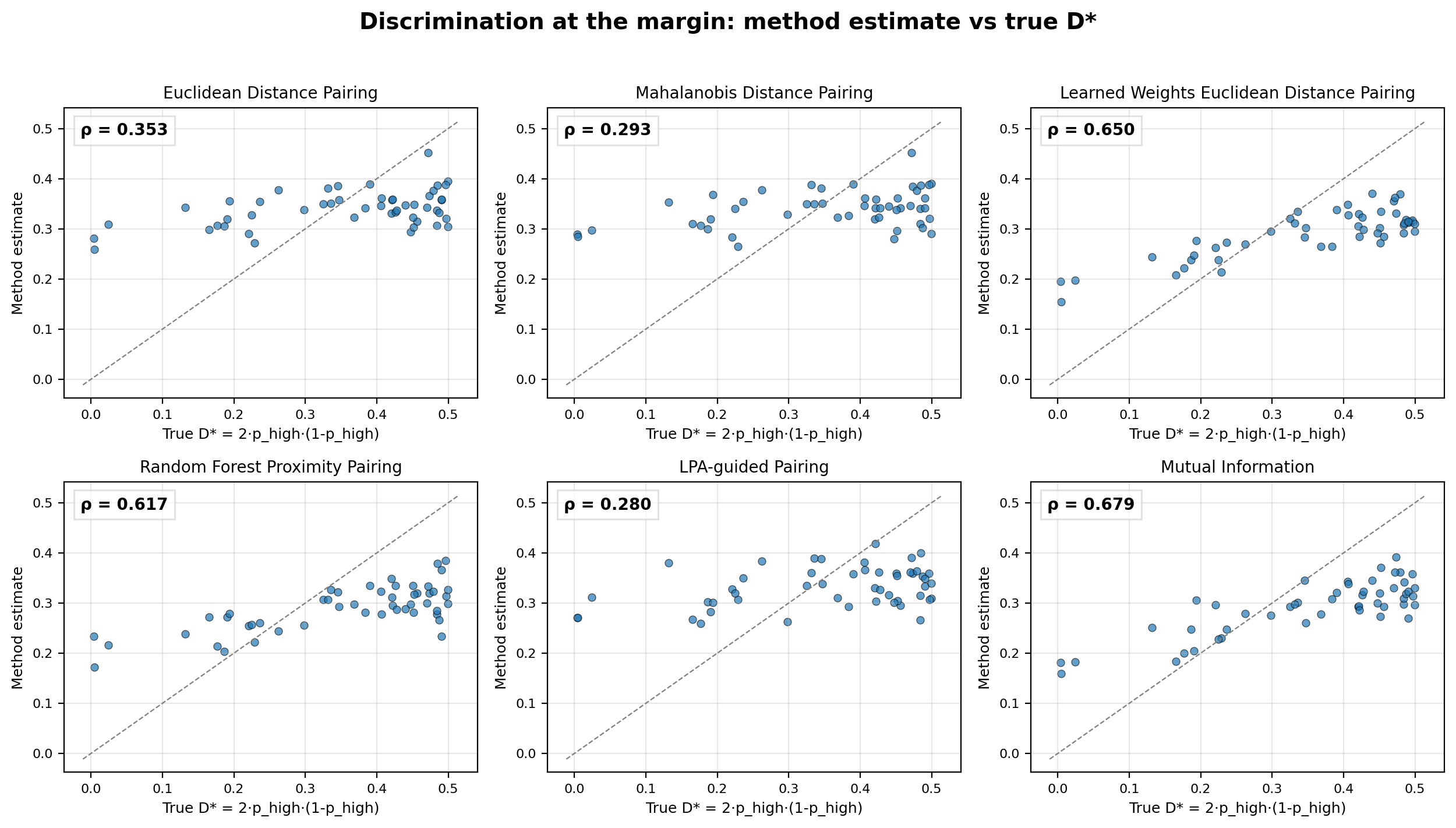}
\caption{\textbf{Continuous-heterogeneity SCORE2 experiment: correlation between prediction and true discordance rate} Each point is one of the $J=50$ physicians. For discordance-rate methods, the dashed diagonal indicates equality between the estimated discordance rate and the true $D^\star_j$. Spearman correlations against the true $D^\star_j$ are printed in the upper-left corner of each panel. The GLMM is analysed separately because its Pearson-residual score is not on the discordance-rate scale.}
\label{fig:scatter_continuous}
\end{figure*}

The continuous-heterogeneity SCORE2 experiment evaluates whether methods discriminate physicians when inconsistency varies along a continuum rather than across five well-separated groups. Figure~\ref{fig:scatter_continuous} plots each per-physician method score against the true discordance $D^\star_j$ and displays the Spearman correlation between each method ranking and the manual pairing.

Rank preservation is substantially weaker under continuous heterogeneity than in the five-group SCORE2 experiment. Supervised feature-weighted methods and the GLMM retain the strongest associations with the true $D^\star_j$: Mutual Information ($\rho=0.679$), Learned Weights ($\rho=0.650$), GLMM ($\rho=0.630$) and RF Proximity ($\rho=0.617$). In contrast, the unsupervised methods show weaker discrimination: Euclidean ($\rho=0.353$), Mahalanobis ($\rho=0.293$) and LPA-guided pairing ($\rho=0.280$). 

The absolute scale of the estimates is also compressed. Figure~\ref{fig:scatter_continuous} shows that discordance-rate methods map the full $[0,0.5]$ range of true $D^\star_j$ to a narrower range of estimated values, with estimates clustering around intermediate discordance levels. This compression is consistent with the positive bias observed for low-heterogeneity physicians in the five-group SCORE2 experiment.

\subsection{Progressive multi-rule experiments results}
\label{sec:results_cross}
The quantile-calibrated threshold design (Section~\ref{sec:gen_progressive}) yields a non-empty eligible stratum in every physician panel of every experiment, so the ground-truth discordance is defined throughout. Table~\ref{tab:cross_experiment} summarises the aggregates for the seven discordance-rate methods; the GLMM is analysed separately because its Pearson-residual score is not on the discordance-rate scale.

\begin{table*}
\centering
\caption{\textbf{Cross-experiment summary.} $\overline{\Dbar}$: mean delta over experiments. $\overline{|\Dbar|}$: mean absolute delta. Med.: median delta. \% Pos.: fraction of experiments with $\Dbar > 0$. Brackets give 95\% percentile bootstrap CIs over the 90 experiments (Section~\ref{sec:methods_overview}). The GLMM is not diplayed here because its Pearson-residual score is not on the discordance-rate scale.}
\label{tab:cross_experiment}
\setlength{\tabcolsep}{3pt}
\resizebox{\textwidth}{!}{%
\begin{tabular}{@{}lcccc@{}}
\toprule
Discordance-rate Method & $\overline{\Dbar}$ & $\overline{|\Dbar|}$ & Med. & \%\,Pos. \\
\midrule
Euclidean    & $+$.063 [$+$.056,\,$+$.069] & .064 [.059,\,.070] & $+$.063 [$+$.059,\,$+$.070] & 93.4 [87.9,\,97.8] \\
Mahalanobis  & $+$.063 [$+$.056,\,$+$.069] & .064 [.059,\,.070] & $+$.063 [$+$.060,\,$+$.070] & 93.4 [87.9,\,97.8] \\
Genetic Mah. & $+$.061 [$+$.054,\,$+$.067] & .063 [.058,\,.069] & $+$.061 [$+$.058,\,$+$.068] & 93.4 [87.9,\,97.8] \\
LPA-guided   & $+$.063 [$+$.057,\,$+$.070] & .065 [.060,\,.071] & $+$.066 [$+$.059,\,$+$.071] & 93.4 [87.9,\,97.8] \\
\textbf{Learned Wts}  & $+$.008 [$+$.001,\,$+$.015] & \textbf{.027 [.023,\,.031]} & $+$.004 [$-$.009,\,$+$.012] & 54.9 [44.0,\,64.8] \\
\textbf{Mutual Info}  & $+$.012 [$+$.005,\,$+$.019] & \textbf{.028 [.023,\,.033]} & $+$.006 [$-$.000,\,$+$.014] & 59.3 [49.5,\,69.2] \\
\textbf{RF Prox.}     & $-$.033 [$-$.037,\,$-$.030] & \textbf{.034 [.031,\,.037]} & $-$.034 [$-$.038,\,$-$.029] &  1.1 [0.0,\,3.3] \\
\bottomrule
\end{tabular}
}
\end{table*}

Two observations emerge. Unsupervised matchers and Genetic Mahalanobis (Euclidean, Mahalanobis, LPA-guided, Genetic Mahalanobis) cluster around $\overline{\Dbar} \approx +0.06$, $\overline{|\Dbar|} \in [0.063, 0.065]$, and overestimate in $ 93\%$ of experiments. LPA-guided is statistically indistinguishable from Euclidean and Mahalanobis: when the eligibility rule lies outside the dominant axis of the latent mixture, the projection adds no information. Supervised feature-weighted methods are markedly more accurate: their $\overline{|\Dbar|}$ confidence intervals (Learning $[.023, .031]$, MI $[.023, .033]$, RF Proximity $[.031, .037]$) do not overlap those of the unsupervised matchers nor of Genetic Mahalanobis ($[.058, .071]$). Learned Weights ($\overline{|\Dbar|}=0.027$) and Mutual Information ($0.028$) overestimate in $54.9\%$ and $59.3\%$ of experiments respectively; RF Proximity ($\overline{|\Dbar|}=0.034$) underestimates in $98.9\%$ of experiments, with $\overline{\Dbar} = -0.033$.

Mechanistically, RF Proximity is trained to predict the prescription outcome; tree splits therefore tend to separate patients with $y=1$ from patients with $y=0$. Cross-outcome pairs share fewer terminal leaves, have lower proximity, and are less likely to be selected as nearest neighbours. This mechanically enriches the matched set in concordant pairs and explains the systematic underestimation observed for RF Proximity.

\subsection{Effect of rule complexity}
\label{sec:results_complexity}
Table~\ref{tab:window_size} reports $\overline{\Dbar}$ stratified by window size.

\begin{table*}[htbp]
\centering
\caption{\textbf{Mean $\Dbar$ by window-size category}, with 95\% percentile bootstrap CIs over experiments within each bin (Section~\ref{sec:methods_overview}). The SCORE2 column ($w=0$) is a single experiment and is shown as a point estimate; its per-physician bootstrap CI is reported in Table~\ref{tab:score2_groups}. The other bins pool 18 ($w=1$), 30 ($w=2$--$3$), 30 ($w=4$--$6$) and 12 ($w=7$--$9$) experiments. The GLMM is not diplayed here because its Pearson-residual score is not on the discordance-rate scale.}
\label{tab:window_size}
\scriptsize
\setlength{\tabcolsep}{3pt}
\begin{tabular}{@{}lccccc@{}}
\toprule
Method & SCORE2$^{\dagger}$ & $w{=}1$ & $w{=}2$--$3$ & $w{=}4$--$6$ & $w{=}7$--$9$ \\
\midrule
Euclidean    & $+$.081 & $+$.036 [$+$.020,\,$+$.050] & $+$.052 [$+$.043,\,$+$.061] & $+$.079 [$+$.069,\,$+$.088] & $+$.086 [$+$.075,\,$+$.095] \\
Mahalanobis  & $+$.081 & $+$.036 [$+$.020,\,$+$.050] & $+$.052 [$+$.043,\,$+$.061] & $+$.079 [$+$.070,\,$+$.088] & $+$.086 [$+$.075,\,$+$.095] \\
Genetic Mah. & $+$.078 & $+$.031 [$+$.015,\,$+$.045] & $+$.051 [$+$.042,\,$+$.059] & $+$.078 [$+$.068,\,$+$.087] & $+$.087 [$+$.078,\,$+$.096] \\
LPA-guided   & $+$.088 & $+$.036 [$+$.019,\,$+$.051] & $+$.053 [$+$.044,\,$+$.062] & $+$.079 [$+$.070,\,$+$.088] & $+$.087 [$+$.076,\,$+$.096] \\
Learned Wts  & $+$.028 & $-$.026 [$-$.034,\,$-$.019] & $-$.009 [$-$.014,\,$-$.004] & $+$.026 [$+$.017,\,$+$.035] & $+$.055 [$+$.047,\,$+$.063] \\
Mutual Info  & $+$.028 & $-$.027 [$-$.033,\,$-$.021] & $-$.005 [$-$.010,\,$-$.000] & $+$.031 [$+$.023,\,$+$.039] & $+$.062 [$+$.051,\,$+$.072] \\
RF Proximity & $+$.029 & $-$.038 [$-$.045,\,$-$.031] & $-$.035 [$-$.040,\,$-$.030] & $-$.031 [$-$.036,\,$-$.026] & $-$.033 [$-$.042,\,$-$.024] \\
\bottomrule
\multicolumn{6}{l}{$^{\dagger}$ Single experiment; see Table~\ref{tab:score2_groups} for the per-physician $\Dbar$ CI.} \\
\end{tabular}
\end{table*}

For single-marker rules ($w=1$), unweighted methods and LPA-guided show a small positive bias ($+0.036$ for Euclidean, Mahalanobis, LPA-guided; $+0.031$ for Genetic Mahalanobis), while feature-weighted methods undershoot ($-0.026$ to $-0.038$). For multi-marker rules (\mbox{$w \geq 4$}), unweighted methods, Genetic Mahalanobis and LPA-guided overestimate increasingly with complexity ($+0.08$ at \mbox{$w{=}4$--$6$}; $+0.09$ at $w{=}7$--$9$). Learned Weights and Mutual Information stay close to zero, with a moderate positive bias only at the largest windows. RF Proximity remains consistently negative ($-0.03$ on average), driven by its tendency to over-cluster patients with the same predicted outcome.

\subsection{Robustness across passes}
\label{sec:results_robustness}
The two passes share the rule structure but use different random thresholds. Pass-level differences are below $0.005$ for every method (Table~\ref{tab:passes}); the patterns reflect rule structure, not specific thresholds. The two passes are statistically indistinguishable: per-method pass-1 and pass-2 CIs overlap entirely (Table~\ref{tab:passes}).

\begin{table}[htbp]
\centering
\caption{Mean $\Dbar$ by pass (progressive-rule experiments only, $n = 45$ per pass), with 95\% percentile bootstrap CIs over experiments (Section~\ref{sec:methods_overview}).}
\label{tab:passes}
\footnotesize
\setlength{\tabcolsep}{3pt}
\begin{tabular}{@{}lcc@{}}
\toprule
Method & Pass~1 & Pass~2 \\
\midrule
Euclidean       & $+$.062 [$+$.052,\,$+$.071] & $+$.063 [$+$.054,\,$+$.072] \\
Mahalanobis     & $+$.062 [$+$.052,\,$+$.071] & $+$.063 [$+$.053,\,$+$.072] \\
Genetic Mah.    & $+$.059 [$+$.049,\,$+$.068] & $+$.062 [$+$.053,\,$+$.072] \\
LPA-guided      & $+$.062 [$+$.051,\,$+$.071] & $+$.064 [$+$.054,\,$+$.073] \\
Learned Wts     & $+$.008 [$-$.002,\,$+$.017] & $+$.008 [$-$.002,\,$+$.018] \\
Mutual Info     & $+$.010 [$+$.000,\,$+$.019] & $+$.014 [$+$.003,\,$+$.024] \\
RF Proximity    & $-$.034 [$-$.039,\,$-$.030] & $-$.034 [$-$.038,\,$-$.029] \\
\bottomrule
\end{tabular}
\end{table}

\subsection{Robustness across cohort and physician panel size}
\label{sec:results_sensitivity}
We replicate SCORE2 on a $4 \times 5$ grid: $n \in \{5\,000, 10\,000, 20\,000, 30\,000\}$ and $j \in \{5, 10, 20, 50, 100\}$, with $B=10$ bootstrap replicates per cell. Six matching methods are recomputed (Euclidean, Mahalanobis, Learned Weights, Mutual Information, RF Proximity, LPA-guided). Full curves are in Appendix~\ref{app:sensitivity_curves} (Figure~\ref{fig:sensitivity_curves}).

The qualitative ordering of methods is preserved across the entire grid. Within each physician panel size, the bias of each method shifts smoothly with $n_{\text{patients}}$. Across the 20 cells and six methods, the most pessimistic bootstrap mean delta is $-0.109$ (Learned Weights, $n_{\text{patients}}=30\,000$, $J=5$); the most pessimistic positive bias is $+0.030$ (Euclidean, $5\,000$, $100$). All other cells satisfy $|\overline{\Dbar}| \leq 0.10$, so the absolute disagreement with the ground truth is bounded by approximately 11 percentage points across the grid.

\subsection{GLMM Pearson-residual overdispersion}
\label{sec:results_glmm}
The GLMM score $\widehat{\mathrm{OD}}_j$ is not directly comparable to discordance rates. On SCORE2, its Spearman correlation with the ground truth is $\rho = 0.971$ ($p < 10^{-4}$), the second-highest across methods after RF Proximity ($\rho = 0.976$). Figure~\ref{fig:score2_results} shows that $\widehat{\mathrm{OD}}_j$ separates the five heterogeneity groups cleanly. The score should still be read as a heuristic ranking statistic, not a calibrated variance (Section~\ref{sec:method_glmm}).

\section{Discussion}
\label{sec:discussion}

This study addresses a methodological question: whether intra-physician prescribing inconsistency can be quantified reliably when the true decision rule is known by construction. We therefore benchmark eight blind discordance-analysis methods against a synthetic ground truth. This framing is deliberately validation-oriented. It does not establish a directly deployable clinical metric, but identifies which analytical strategies remain informative under controlled variations in rule complexity, cohort structure and physician heterogeneity.

\subsection{Principal findings}
The 94 experiments yield five conclusions. (i) On the 5-group SCORE2 experiment, every method preserves the physician ranking with high fidelity (Spearman $\rho \geq 0.89$). Under a continuous-heterogeneity physician model, rank preservation drops to $\rho \in [0.28, 0.68]$: but the supervised feature-weighted methods (Mutual Information, Learned Weights, RF Proximity) and the GLMM retain a useful signal. Rank preservation, the property most relevant to quality-improvement trainings, is therefore strong but conditional on physicians not being too close to each other. (ii) Supervised feature-weighted methods are the most accurate in absolute value: Learned Weights ($\overline{|\Dbar|} = 0.027$), Mutual Information ($0.028$) and RF Proximity ($0.034$) remain below $4\%$ mean absolute error across the experiments. (iii) No single method dominates across rule structures; the choice should be guided by the expected complexity of the prescription rule. (iv) While unweighted distances (Euclidean, Mahalanobis) and LPA-guided pairing are reasonable defaults for single-marker rules, feature-weighted methods are preferable, especially for complex multi-marker rules.

\subsection{Quantile-calibrated thresholds and the dilution effect}
\label{sec:discussion_underestimation}
Under uniform-in-range threshold sampling, thresholds are sampled uniformly within each covariate's empirical range. The marginal eligibility probability shrinks as $\sim 0.5^w$, leaving the eligible region empty for large $w$. Two pathologies follow: experiments with undefined ground truth are discarded; blind methods underestimate, because most matched pairs join two non-eligible patients (both with prescription probability $p^{\text{low}}$), diluting the discordance below its eligible-stratum value. The quantile-calibrated rule with $p^\star \sim \mathcal{U}(0.2, 0.8)$ eliminates both pathologies: every condition has a valid ground truth, and most methods now exhibit a mild positive bias. RF Proximity is the only systematic underestimator. This is not a residual dilution effect but a consequence of how RF proximity is constructed. The random forest is trained to predict the prescription outcome $y$; tree splits therefore tend to separate patients with $y=1$ from patients with $y=0$. Cross-outcome patient pairs share fewer terminal leaves, have lower proximity, and are less likely to be selected as nearest neighbours. The matched set is consequently enriched in concordant pairs, which mechanically compresses the estimated discordance rate downward.

The high inter-method correlations in the 5-group SCORE2 experiment should be interpreted in light of the dominant between-group signal: physicians are placed on five well-separated behavioural rungs, which makes rank preservation relatively easy even for methods that differ substantially in absolute calibration. The continuous-heterogeneity experiment provides a more stringent discrimination test and reveals larger differences between methods.

\subsection{Calibration versus discrimination}
\label{sec:discussion_calibration_discrimination}

The contrast between the five-group and the continuous heterogeneity SCORE2 experiments clarifies the role of rank preservation. In the five-group experiment, physicians are placed on five discrete rungs of $p^{\text{high}}$, creating a strong between-group signal. This makes the rank ordering relatively easy to recover: methods can preserve the physician ranking even when their absolute discordance estimates are biased. The high inter-method correlations observed in this setting should therefore not be interpreted as evidence that the methods are interchangeable, but as evidence that the benchmark contains a dominant separable physician-level signal.

The continuous-heterogeneity experiment is more stringent. When physicians vary along a continuum, small errors in patient matching or feature weighting are more likely to change the physician ordering. The observed drop in Spearman correlation therefore tempers the conclusion that all methods preserve rankings with high fidelity. In this setting, the deployment-relevant signal is retained mainly by supervised feature-weighted methods and by the GLMM. This supports Mutual Information weighting, Learned Weights, RF Proximity and the GLMM as the most relevant candidates for physician profiling on observational data, while suggesting that unweighted distances and LPA-guided pairing are less suitable when the primary goal is rank discrimination.

\subsection{Entropy of the prescribing process versus physician inconsistency}
\label{sec:discussion_entropy_vs_inconsistency}
Every method estimates the entropy of the prescribing process: the probability that two equivalent prescribing situations yield discordant decisions, conditional on the available information. This entropy is not the intrinsic inconsistency of the physician.

In the synthetic framework, the two notions coincide by construction: the simulated physician is a Bernoulli sampler with no other source of stochasticity. In observational data, the same empirical entropy decomposes into at least three components: genuine reasoning-level inconsistency; legitimate stochasticity from patient or contextual factors not captured by the observed covariates (e.g.\ preferences, contraindications, longitudinal context); and measurement and documentation noise. None of these components is identifiable from prescriptions alone. The proposed methods, when applied to real data, should therefore be read as estimators of the prescribing entropy, meaning an upper bound on, or proxy for, intrinsic inconsistency. Closing this gap requires richer covariates or designs where all covariates are controlled.

\subsection{Limitations}
\textbf{Synthetic data simplify reality.} In the main benchmark, covariates are sampled independently from truncated normal or Bernoulli marginals; prescribing rules are deterministic functions of observable covariates; all physicians share the same eligibility rule. Two SCORE2 sensitivity analyses (Appendix~\ref{app:sensitivity_data_generation}) probe departures from covariate independence (Gaussian-copula correlation) and from Gaussian marginals (lognormal HbA1c).

\textbf{Many real-world determinants are absent.} Patient preferences, contraindications and drug interactions, incomplete documentation, contextual clinical judgement, and system-level constraints (formulary, reimbursement, local guidelines) are not represented. The fraction of apparent inconsistency attributable to such legitimate sources cannot be quantified here.

\textbf{Comparability is reduced to a pairwise score.} Real comparability is nonlinear, hierarchical and context-dependent (e.g.\ thresholds shift with diabetes status; contraindications act as gating criteria). None of the methods reproduces this fully.

\textbf{Marginal distributions} are largely symmetric apart from HbA1c in the sensitivity analysis.

\textbf{Physician-panel size may affect rank discrimination.} The continuous-heterogeneity SCORE2 benchmark uses $J=50$ physicians with an average of $400$ patients per physician. This panel size is informative for validating method behaviour under a continuous physician model, but it may be larger than panels available in some real-world profiling settings. Rank correlations observed in this experiment should therefore be interpreted as deployment-oriented but not definitive evidence of small-panel performance. Additional experiments with more physicians and fewer patients per physician are needed to assess whether the ranking signal of supervised methods and the GLMM is preserved in sparse physician panels.

\textbf{Computational cost.} Genetic Mahalanobis takes ${\sim}10$~min per experiment for $n=10,000$, $p=9$. Given its computational cost and its near-identical performance to Euclidean and Mahalanobis matching across the benchmark, Genetic Mahalanobis is not recommended as a default method in this setting.

\textbf{The GLMM score is a Pearson-residual heuristic.} Bernoulli outcomes do not admit overdispersion in the strict statistical sense; a calibrated variance would require additional structural assumptions (e.g.\ informative prior on an observation-level random effect; hierarchical Bernoulli--Beta). 

\textbf{Single binary outcome.} Real physicians make multiple interacting prescribing decisions simultaneously.

\textbf{Observable-only assumption.} The framework presumes that all features informing eligibility are observed; in practice, free-text history, physical-examination findings and patient-reported preferences often are not.

\textbf{Unobservable real-world ground truth.} In observational data, the closed-form ground truth used here is not accessible.

\section{Conclusion}
\label{sec:conclusion}
This framework of eight blind discordance-analysis methods quantifies the intra-physician variability closely to the ground truth, in 94 synthetic experimental conditions. The framework combines two SCORE2 cardiovascular risk scenario with two sensitivity and 90 progressive multi-rule experiments under a quantile-calibrated threshold design. To our knowledge, this is the first systematic and controlled comparison of methods for intra-physician variability quantification.

In the 5-group SCORE2 benchmark, all methods preserve the physician rank ordering with high fidelity (Spearman $\rho \geq 0.89$). However, under continuous physician heterogeneity, rank preservation weakens substantially for unsupervised methods and is retained mainly by supervised feature-weighted methods and the GLMM. These results support Mutual Information weighting, Learned Weights, RF Proximity and the GLMM as the most relevant candidates for real-world physician profiling.

Feature-weighted approaches (Learned Weights, Mutual Information, RF Proximity) achieve the lowest mean absolute error ($\overline{|\Dbar|} \leq 0.034$), outperforming unweighted metrics on complex prescription rules. SCORE2 sensitivity analyses (Appendix~\ref{app:sensitivity_data_generation}) confirm the conclusions under covariate correlation and non-Gaussian marginals.

This work is a proof-of-concept methodological framework, not evidence for clinical deployment. Future work should validate these methods on observational prescribing data, develop strategies to separate legitimate variation from genuine inconsistency, and extend the framework to multiple interacting prescribing decisions.


\bigskip

\appendix

\section{Distribution of Pearson residuals from the GLMM\label{app1}}
\label{app:pearson_residuals}
\vspace*{12pt}

\begin{figure}[htbp]
\centering
\includegraphics[width=\linewidth]{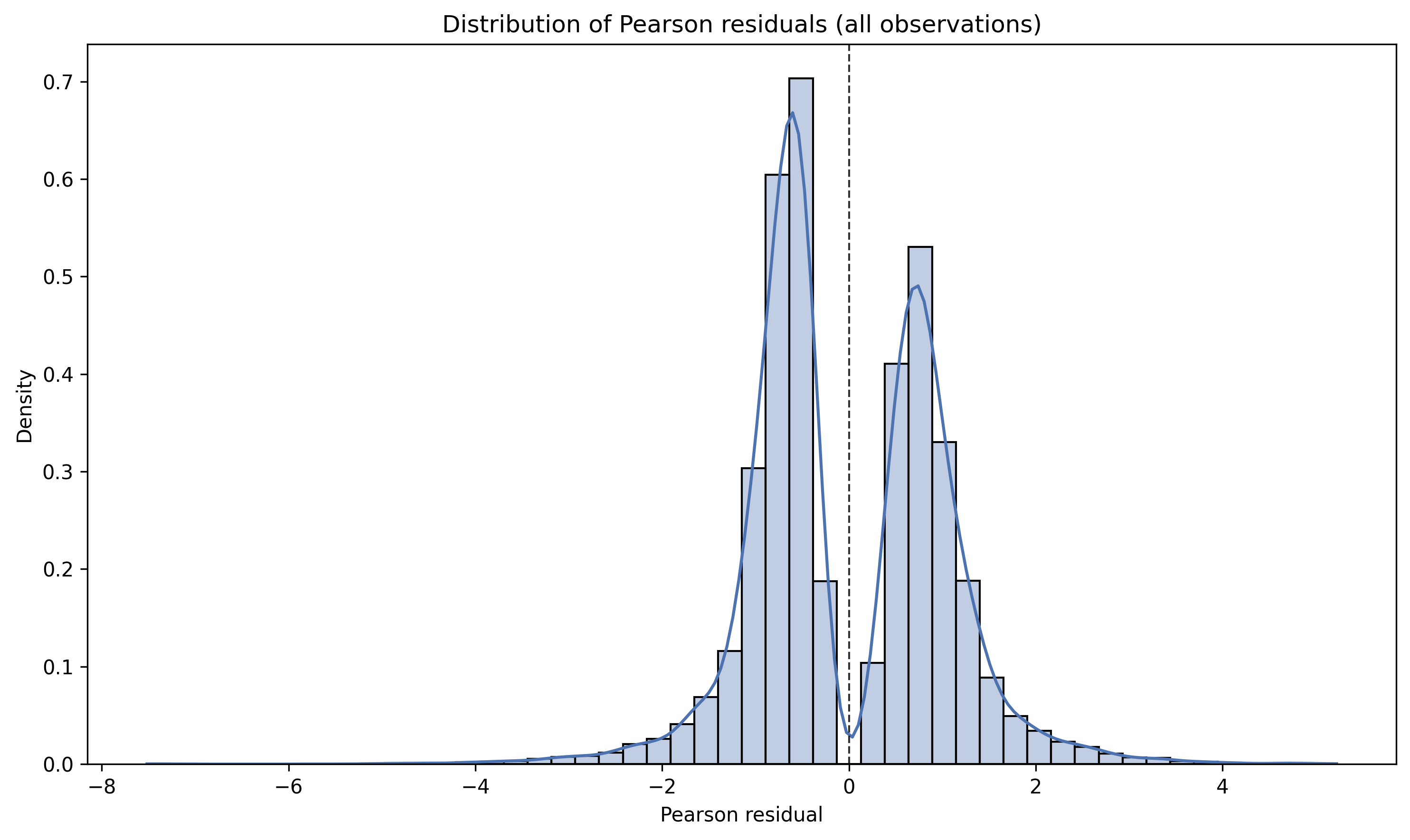}
\caption{Pearson residuals from the GLMM (SCORE2 experiment). Residuals are centred near zero; the spread reflects the unexplained patient-level variability that drives $\widehat{\mathrm{OD}}_j$.}
\label{fig:pearson_residuals_distribution}
\end{figure}

The distribution of Pearson residuals is normal, and plotted in Fig~\ref{fig:pearson_residuals_distribution}

\section{Calibration curve of the GLMM}
\label{app:calibration_curve}

\begin{figure}[htbp]
\centering
\includegraphics[width=\linewidth]{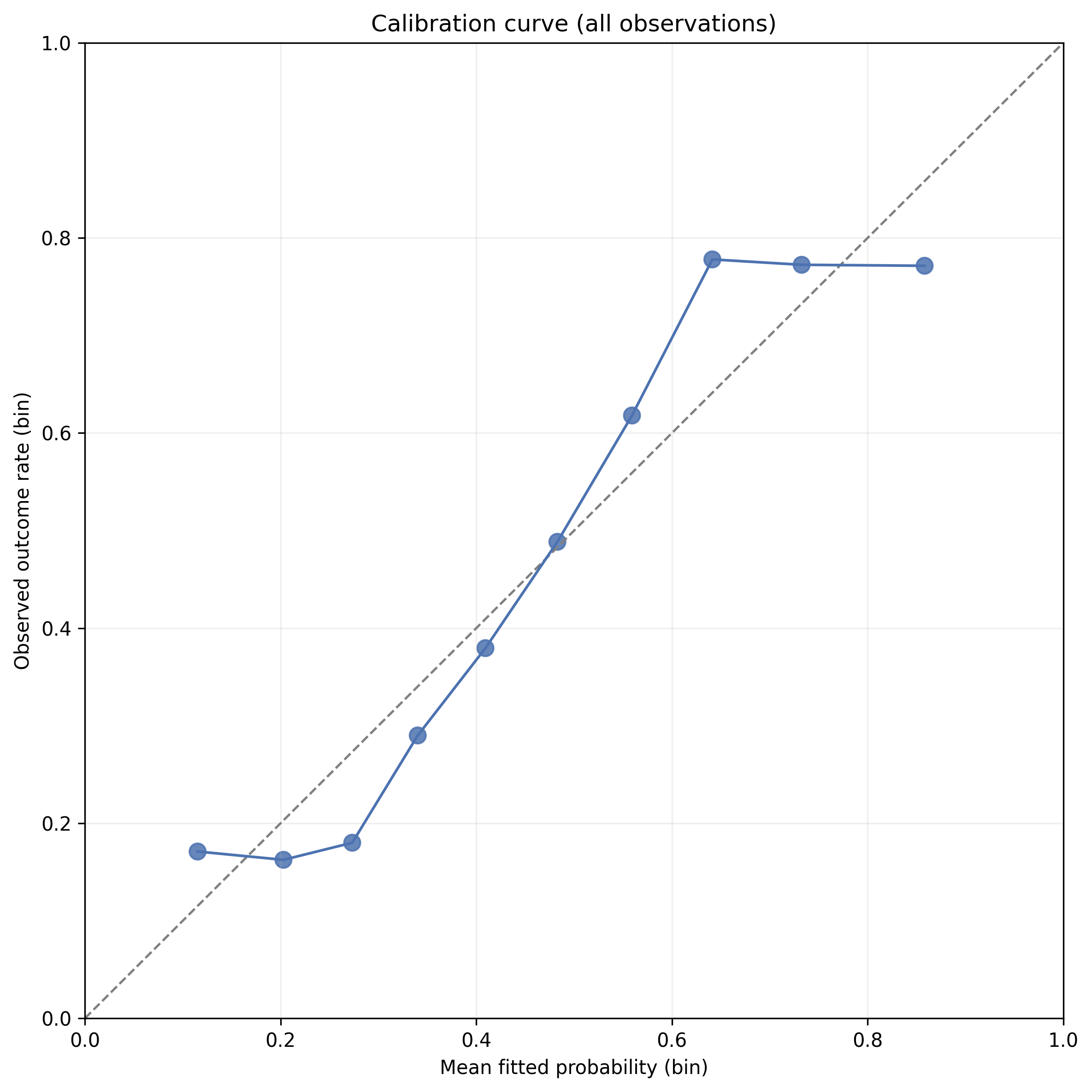}
\caption{Calibration curve of the GLMM (SCORE2 experiment). Observed prescription frequencies versus predicted probabilities.}
\label{fig:calibration_curve}
\end{figure}

The calibration curve of the GLMM for the SCORE2 experiment, shown in Fig.~\ref{fig:calibration_curve}, indicates that the mean fitted probabilities closely match the observed outcomes, supporting the hypothesis that Pearson residuals capture the physician effect.

\section{Sensitivity to cohort and physician panel size}
\label{app:sensitivity_curves}

\begin{figure*}[htbp]
\centering
\includegraphics[width=\linewidth]{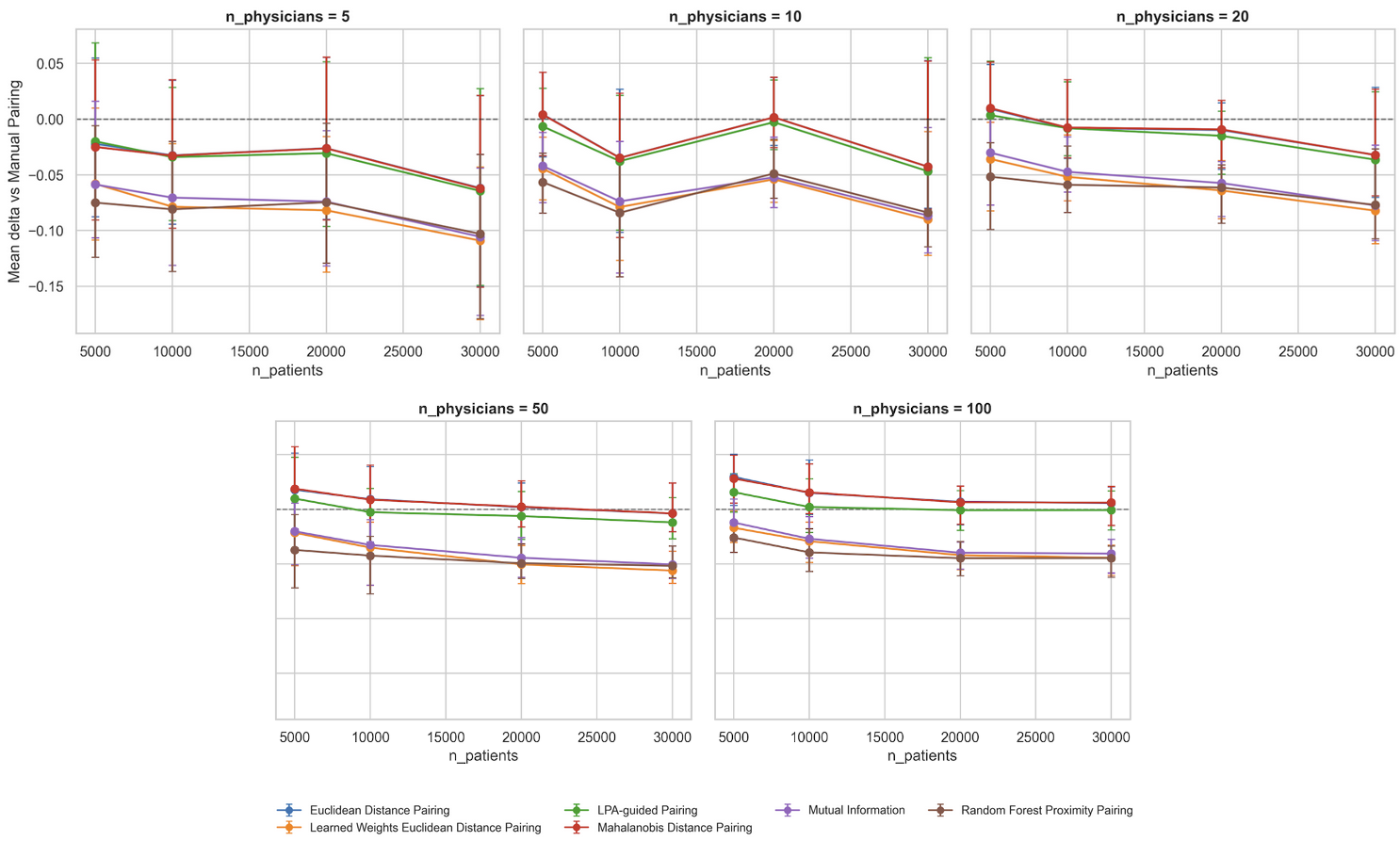}
\caption{\textbf{Bootstrap mean delta versus manual pairing across cohort and physician panel size (SCORE2 experiment).} \\Sub-panels correspond to $n_{\text{physicians}} \in \{5, 10, 20, 50, 100\}$; abscissa is $n_{\text{patients}} \in \{5\,000, 10\,000, 20\,000, 30\,000\}$; ordinate is the bootstrap mean delta ($B=10$ replicates per cell). Vertical bars are $95\%$ bootstrap CIs. Six matching methods are reported (Euclidean, Mahalanobis, Learned Weights, Mutual Information, RF Proximity, LPA-guided). The worst-case mean delta across the grid is $\approx 0.11$; all other cells satisfy $|\overline{\Dbar}| \leq 0.10$.}
\label{fig:sensitivity_curves}
\end{figure*}

The methods sensitivity to cohort and physician panel size is explored in Fig~\ref{fig:sensitivity_curves}.

\section{Sensitivity to covariate correlation and marginal distribution}
\label{app:sensitivity_data_generation}

We probe two structural departures from the main experiment. Each sensitivity run uses an independently sampled cohort drawn under the modified data-generating process; the physician groups (Table~\ref{tab:physician_groups}), the SCORE2 rule (Section~\ref{sec:gen_prescribing}), and the pairing and caliper procedures (Sections~\ref{sec:method_nn} and \ref{sec:method_genmah}) are unchanged. Numerical comparisons against the main-benchmark SCORE2 cohort (Table~\ref{tab:score2_groups}) should be interpreted up to Monte-Carlo noise from the different cohort seed. Configurations are provided in the \href{https://github.com/alaedinebe/physician-intra-variability}{repository}.

\subsection{Gaussian-copula correlation between non-HDL and LDL cholesterol}

\textbf{Setup.} A Gaussian copula imposes a target Spearman correlation $\rho_{\text{copula}} = 0.8$ between non-HDL and LDL cholesterol while preserving the marginals; the realised correlation on the 10\,000-patient cohort is $\hat\rho = 0.786$. LDL and non-HDL cholesterol are physiologically coupled in real cohorts ($\rho \in [0.7, 0.9]$ in observational European data). The other seven covariates remain independent.

\textbf{Results.} Table~\ref{tab:sensitivity_copula} reports the per-group discordance rates and $\Dbar$. Every method preserves the physician ranking ($\rho \geq 0.94$ versus the ground truth, with Mutual Information at $\rho = 0.97$). The ordering of methods by absolute accuracy is unchanged: feature-weighted methods (Learned Weights, RF Proximity, Mutual Information) remain the most accurate; unsupervised distance methods and Genetic Mahalanobis cluster at $\Dbar \in [+0.06, +0.07]$. The absolute biases of the unsupervised matchers and LPA-guided pairing fall slightly below their main-benchmark SCORE2 values, consistent with the latent Gaussian-mixture space absorbing part of the non-HDL/LDL coupling.

\begin{table*}[htbp]
\centering
\caption{\textbf{Sensitivity B.1.} SCORE2 with Gaussian-copula correlation $\rho_{\text{copula}}=0.8$ between non-HDL and LDL cholesterol (realised $\hat\rho=0.786$ on the 10\,000-patient cohort). Last column: Spearman correlation with the ground truth.}
\label{tab:sensitivity_copula}
\footnotesize
\begin{tabular}{@{}lccccccc@{}}
\toprule
Method & Grp\,1 & Grp\,2 & Grp\,3 & Grp\,4 & Grp\,5 & $\Dbar$ & $\rho_{\text{GT}}$ \\
\midrule
Ground truth (Manual) & 0.000 & 0.189 & 0.320 & 0.429 & 0.500 & --- & 1.00 \\
Euclidean             & 0.217 & 0.278 & 0.357 & 0.425 & 0.493 & $+$0.066 & 0.95 \\
Mahalanobis           & 0.217 & 0.302 & 0.347 & 0.417 & 0.496 & $+$0.068 & 0.95 \\
Learned Wts           & 0.127 & 0.232 & 0.308 & 0.411 & 0.490 & $+$0.026 & 0.96 \\
Genetic Mah.          & 0.213 & 0.292 & 0.347 & 0.424 & 0.510 & $+$0.070 & 0.94 \\
RF Proximity          & 0.143 & 0.228 & 0.317 & 0.380 & 0.498 & $+$0.025 & 0.96 \\
\textbf{LPA-guided}   & \textbf{0.223} & \textbf{0.281} & \textbf{0.339} & \textbf{0.424} & \textbf{0.496} & $\mathbf{+}$\textbf{0.065} & \textbf{0.94} \\
Mutual Info           & 0.150 & 0.249 & 0.323 & 0.414 & 0.501 & $+$0.040 & 0.97 \\
GLMM ($\widehat{\mathrm{OD}}_j$) & 0.54 & 0.73 & 0.93 & 1.22 & 1.92 & --- & 0.95 \\
\bottomrule
\end{tabular}
\end{table*}

\begin{figure*}[htbp]
\centering
\includegraphics[width=\linewidth]{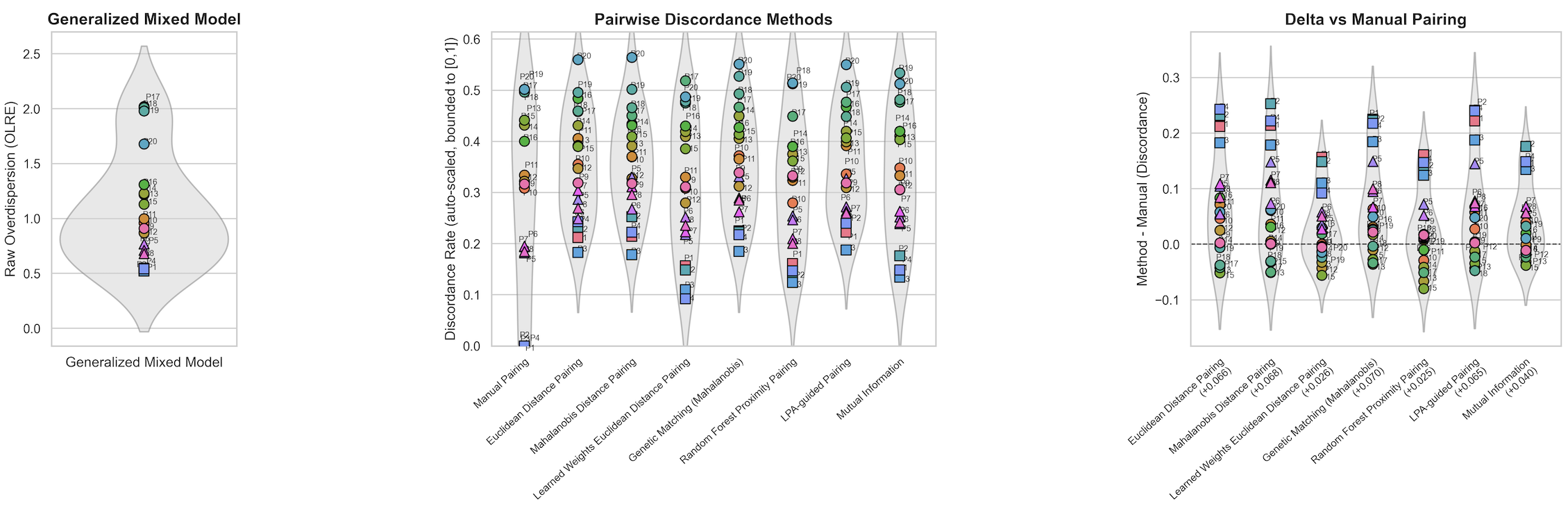}
\caption{\textbf{Sensitivity B.1.} SCORE2 with Gaussian-copula correlation $\rho_{\text{copula}}=0.8$ between non-HDL and LDL cholesterol.}
\label{fig:sensitivity_copula}
\end{figure*}

\subsection{Lognormal HbA1c marginal}

\textbf{Setup.} HbA1c is drawn from a right-skewed lognormal with parameters $\mu = 1.846$, $\sigma = 0.228$ (calibrated, before clipping, to $\mathbb{E}[X] \approx 6.5\%$ and $\text{SD}[X] \approx 1.5\%$), then clipped to $[4, 12]\%$. The realised marginal skewness on the 10\,000-patient cohort is $\hat g_1 = 0.70$ (mean $6.48\%$, SD $1.48\%$). The eight other covariates retain their main-benchmark marginals.

\textbf{Results.} Table~\ref{tab:sensitivity_lognormal}. Rank preservation is maintained ($\rho \geq 0.94$). The absolute biases of the unsupervised matchers and LPA-guided pairing are essentially unchanged from the independent-normal baseline; the supervised feature-weighted methods are slightly improved (RF Proximity: $+0.022$; Mutual Information: $+0.028$). RF Proximity reaches $\rho = 0.99$ with the ground truth. The robust-scaling pre-processing of LPA-guided pairing absorbs the heavy tail without destabilising the method.

\begin{table*}[htbp]
\centering
\caption{\textbf{Sensitivity B.2.} SCORE2 with right-skewed lognormal HbA1c (realised skewness $\hat g_1 = 0.70$ on the 10\,000-patient cohort; mean $6.48\%$, SD $1.48\%$). Last column: Spearman correlation with the ground truth.}
\label{tab:sensitivity_lognormal}
\footnotesize
\begin{tabular}{@{}lccccccc@{}}
\toprule
Method & Grp\,1 & Grp\,2 & Grp\,3 & Grp\,4 & Grp\,5 & $\Dbar$ & $\rho_{\text{GT}}$ \\
\midrule
Ground truth (Manual) & 0.000 & 0.178 & 0.306 & 0.446 & 0.502 & --- & 1.00 \\
Euclidean             & 0.205 & 0.320 & 0.340 & 0.458 & 0.508 & $+$0.080 & 0.95 \\
Mahalanobis           & 0.210 & 0.318 & 0.339 & 0.457 & 0.515 & $+$0.081 & 0.95 \\
Learned Wts           & 0.144 & 0.239 & 0.292 & 0.432 & 0.513 & $+$0.038 & 0.97 \\
Genetic Mah.          & 0.211 & 0.310 & 0.361 & 0.437 & 0.525 & $+$0.083 & 0.96 \\
RF Proximity          & 0.119 & 0.217 & 0.303 & 0.410 & 0.492 & $+$0.022 & 0.99 \\
\textbf{LPA-guided}   & \textbf{0.206} & \textbf{0.318} & \textbf{0.343} & \textbf{0.461} & \textbf{0.510} & $\mathbf{+}$\textbf{0.081} & \textbf{0.94} \\
Mutual Info           & 0.127 & 0.232 & 0.295 & 0.426 & 0.493 & $+$0.028 & 0.97 \\
GLMM ($\widehat{\mathrm{OD}}_j$) & 0.55 & 0.74 & 0.89 & 1.17 & 1.99 & --- & 0.98 \\
\bottomrule
\end{tabular}
\end{table*}

\begin{figure*}[htbp]
\centering
\includegraphics[width=\linewidth]{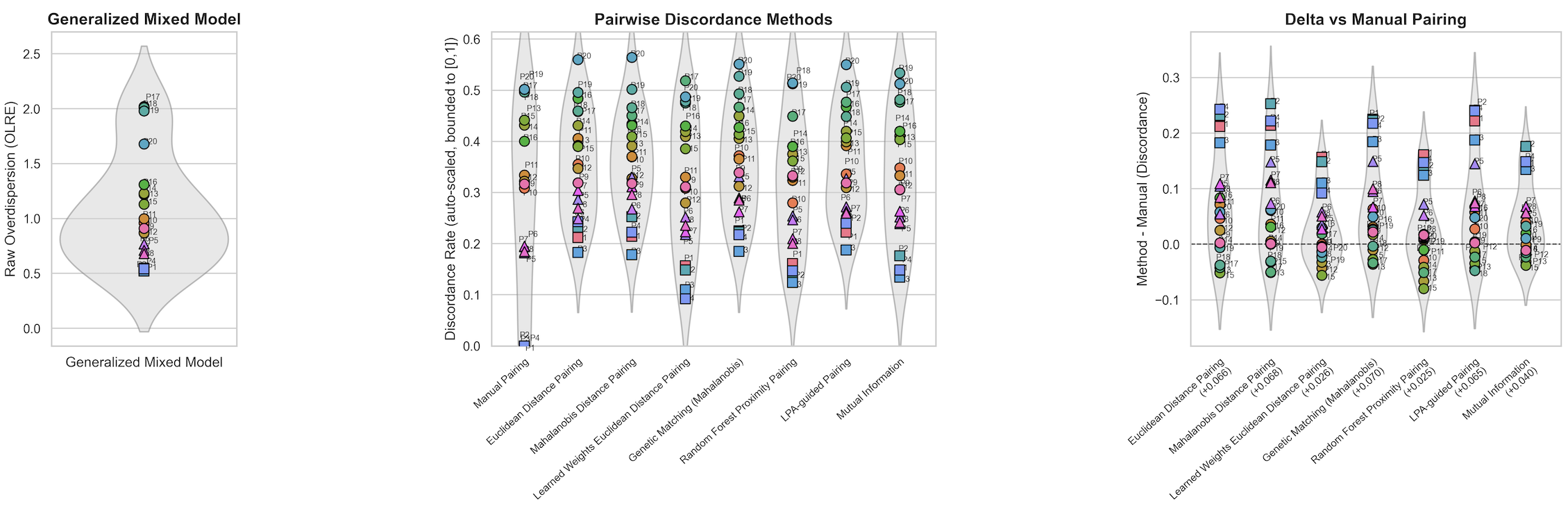}
\caption{\textbf{Sensitivity B.2.} SCORE2 with right-skewed lognormal HbA1c marginal.}
\label{fig:sensitivity_lognormal}
\end{figure*}

\subsection{Synthesis}

Both sensitivity analyses preserve the principal conclusions of the main benchmark. Rank preservation holds ($\rho \geq 0.94$). The ordering by absolute accuracy is unchanged: supervised feature-weighted methods (Learned Weights, Mutual Information, RF Proximity) remain the most accurate, and LPA-guided pairing remains comparable to the unsupervised distance metrics. No method exhibits a destabilising regime change. The recommendations of Section~\ref{sec:discussion} therefore carry over to these two configurations.

\end{document}